\documentclass[a4paper,11pt]{article}
\pdfoutput=1 

\usepackage{jheppub1} 

\usepackage[T1]{fontenc} 

\usepackage{amsmath,amsthm,amssymb,mathrsfs,amsfonts,graphicx,subfigure,tikz,mathtools}

\usepackage{comment}
\usepackage{enumerate}
\usepackage[normalem]{ulem}

\def\bea{\begin{eqnarray}}
\def\eea{\end{eqnarray}}

\def\ba{\begin{array}}
\def\ea{\end{array}}

\def\Tr{\text{Tr}}

\def\J{\mathcal{J}}

\def\C{\mathcal{C}}
\def\V{\mathcal{V}}

\def\O{\mathcal{O}}
\def\T{\mathcal{T}}
\def\D{\mathcal{D}}
\def\M{\mathcal{M}}
\def\N{\mathcal{N}}
\def\L{\mathcal{L}}
\def\H{\mathcal{H}}

\def\Pf{\text{Pf}}

\def\t{\tau}
\def\bra{\langle}
\def\ket{\rangle}
\def\a{\alpha}
\def\b{\beta}

\newcommand{\hc}{\dagger}
\newcommand{\tr}{\mathrm{Tr}}
\newcommand{\vphi}{\varphi}

\newcommand{\Z}{\mathcal{Z}}
\newcommand{\U}{\mathcal{U}}

\title{Effective Field Theory of Operator Scrambling from Strong-to-Weak Symmetry Breaking}

\author[a]{Bai-Lin Cheng,}
\emailAdd{bailincheng@stu.pku.edu.cn}

\author[b]{Shao-Kai Jian}%
\emailAdd{sjian@tulane.edu}
\author[a,c]{and Zhi-Cheng Yang}
\emailAdd{zcyang19@pku.edu.cn}

\affiliation[a]{School of Physics, Peking University, Beijing 100871, China}
\affiliation[b]{Department of Physics and Engineering Physics, Tulane University, New Orleans, Louisiana 70118, USA}
\affiliation[c]{Center for High Energy Physics, Peking University, Beijing 100871, China}

\abstract{Operator scrambling is commonly diagnosed by the growth of out-of-time-ordered correlators (OTOCs), yet a general symmetry principle underlying their effective dynamics has remained elusive. For Brownian or short-time-correlated large-$N$ Majorana systems, we develop a symmetry-based effective field theory for operator scrambling, organized by a strong-to-weak U(1) symmetry breaking in operator space. The key observation is that, in the noninteracting fermion limit, the four-fold Keldysh contour representation of an OTOC admits an emergent strong U(1) symmetry in a doubled Hilbert-space description, even when the original system has no ordinary conserved quantity. The associated slow mode is the phase of the strong-charge creation operator, whose conjugate density is identified with the local operator size. Generic interactions explicitly break the strong symmetry and generate a mass term at lowest order for the would-be Goldstone mode, thereby converting diffusive operator spreading into chaotic growth. We further show that higher-order symmetry breaking terms are tightly constrained by an emergent duality that combines time reversal with contour permutation. This duality fixes the effective action up to quadratic order in the response field, relates the multiplicative noise strength directly to the Lyapunov exponent, and makes the positivity of the Lyapunov exponent a consequence of convergence of the real-time path integral. The resulting OTOC dynamics is governed by a noisy Fisher-Kolmogorov-Petrovsky-Piskunov equation, which captures within a unified framework the early-time exponential growth, ballistic propagation, nonlinear saturation, and stochastic front broadening of operator scrambling. We verify this construction in a Brownian Sachdev-Ye-Kitaev chain, where a direct saddle-point expansion reproduces the symmetry-based effective action. Our results reveal a symmetry origin of operator-size hydrodynamics and provide a principled route to effective theories of quantum scrambling beyond model-specific master equations.}

\begin{document} 
\maketitle
\flushbottom

\newpage

\section{Introduction}

Understanding how quantum information spreads in an isolated many-body system is a central problem in modern quantum dynamics. Under Heisenberg evolution, an initially local operator becomes progressively delocalized in operator space, acquiring weight on operator strings with increasing spatial support and complexity. This process, usually referred to as operator scrambling, provides a microscopic characterization of quantum chaos beyond conventional probes based on spectra or transport. It is closely connected to thermalization~\cite{deutsch1991quantum,srednicki1994chaos,rigol2008thermalization,d2016quantum,liu2014entanglement}, entanglement growth~\cite{nahum2017quantum,nahum2018operator,jonay2018coarse,zhou2020entanglement}, the emergence of hydrodynamic behavior~\cite{rakovszky2018diffusive,von2018operator,PhysRevX.8.031057}, and information-theoretic distinctions between integrable and chaotic dynamics~\cite{alba2019operator,parker2019universal,dowling2023scrambling,xu2024scrambling}.

A particularly useful diagnostic of operator scrambling is the out-of-time-ordered correlator (OTOC)~\cite{sahu2020information,larkin1969quasiclassical,shenker2014black,kitaev2015simple,maldacena2016bound,garttner2017measuring,mi2021information,kitaev2018soft,lin2018out,braumuller2022probing}. Consider, for example, a fermionic system. For two initially separated local fermionic operators $W_0$ and $V_x$, the OTOC measures the growth of their anticommutator under Heisenberg time evolution:
\begin{equation}
    C(x,t)=\frac{1}{\D}{\rm Tr}\left[\{W_0(t),V_x\}^\dagger \{W_0(t),V_x\}\right],
\end{equation}
where $\D$ is the Hilbert-space dimension, so that the trace represents an infinite-temperature expectation value. For bosonic operators or spins, the anticommutator is replaced by a commutator. This squared norm of the anticommutator is directly related to the operator size of $W_0(t)$. At $t=0$, the operator $W_0$ is supported only near the origin and therefore has a vanishing anticommutator with a fermionic operator $V_x$ located far away. Under time evolution, $W_0(t)$ acquires support over an increasingly large spatial region and eventually reaches the vicinity of $x$. Once this occurs, its anticommutator with $V_x$ becomes nonzero and the OTOC begins to grow. In this sense, the OTOC directly probes the spatial growth of operator size.

In chaotic systems, $C(x,t)$ typically grows rapidly near the leading edge of the operator front. In large-$N$ systems, this growth often takes the form~\cite{maldacena2016bound,gu2017local,parker2019universal}
\begin{equation}
    C(x,t)\sim e^{\lambda_L(t-x/v_B)} \,,
\end{equation}
where $\lambda_L$ is the Lyapunov exponent and $v_B$ is the butterfly velocity. The OTOC therefore captures two central aspects of scrambling: how rapidly an initially simple operator grows in operator space, and how this growth propagates across the system. At later times, however, the local OTOC cannot grow indefinitely. Once the evolving operator has become sufficiently scrambled within a local region, its local weight approaches a saturated value. A complete theory of scrambling should therefore capture not only the leading-edge growth and ballistic propagation of the operator front, but also the nonlinear saturation behind the front and the fluctuations of the front itself~\cite{qi2019quantum,gu2022two,zhang2023operator}.

Several complementary approaches to operator scrambling have been developed. Random unitary circuits provide analytically tractable models in which operator fronts propagate ballistically and broaden diffusively~\cite{nahum2018operator,von2018operator,sahu2020information}. They also led to the notion of operator hydrodynamics, even in systems without energy conservation or other ordinary conserved quantities. In systems with conserved charges, the interplay between ballistic operator growth and diffusive hydrodynamic modes produces characteristic long-time tails in OTOCs~\cite{PhysRevX.8.031057,rakovszky2018diffusive}. In another direction, large-$N$ and Brownian models give rise to kinetic or master equations for operator size~\cite{aleiner2016microscopic,xu2019locality,chen2019quantum,zhou2020operator,zhou2023hydrodynamic,agarwal2022emergent,yao2024notes,xu2025dynamics,sunderhauf2019quantum}. These equations often take the form of Fisher-Kolmogorov-Petrovsky-Piskunov (FKPP) equations~\cite{Fisher1937,KPP1937}, or noisy variants thereof~\cite{brunet2006phenomenological}, and naturally describe exponential growth, traveling fronts, saturation, and front fluctuations. Closely related ideas also appear in Sachdev-Ye-Kitaev (SYK) models~\cite{sachdev1992gapless,kitaev2015simple,maldacena2016remarks,polchinski2016spectrum,jian2021note,chen2020many,saad2018semiclassical}, where operator-size distributions provide a refined probe of scrambling~\cite{qi2019quantum,roberts2018operator}. In a complementary direction, effective theories formulated in terms of collective scramblon modes have been developed to describe OTOC growth and saturation in SYK and related chaotic systems~\cite{keldysh1965diagram,gu2022two,zhang2023operator,stanford2024scramblon,stanford2022subleading,choi2023effective}.

These developments suggest that the dynamics of OTOCs admits a hydrodynamic or reaction-diffusion description. However, most existing derivations rely on special microscopic structures: Haar-random circuits, Brownian Hamiltonians, large-$N$ factorization, kinetic equations, or effective classical master equations for operator strings. This raises a more basic question: Is there a symmetry principle behind the effective dynamics of operator scrambling, in the same sense that ordinary hydrodynamics follows from conservation laws and Schwinger-Keldysh consistency conditions~\cite{glorioso2016second,crossley2017effective,liu2018lectures,baggioli2023u,gao2023effective}?

In this work, we answer this question affirmatively. We show that operator scrambling can be organized by a symmetry principle that is, at first sight, rather surprising: a U(1) strong-to-weak spontaneous symmetry breaking in operator space~\cite{PRXQuantum.4.030317,sala2024spontaneous,lessa2025strong,huang2025hydrodynamics,3g6d-gn7b,chen2025strong,hauser2026strong}. This perspective is unusual because the systems of interest need not possess any ordinary conserved quantity. In particular, the quantum dynamics under consideration need not conserve energy or particle number. Instead, the relevant slow mode is tied to the structure of operator evolution itself.

More specifically, the class of models discussed in this work obeys the following assumptions:
\begin{enumerate}
    \item Systems with $N$ Majorana flavors per site, where $N$ is large but finite, and arbitrary even-$q$-body inter- and intra-site couplings.
    \item The coupling $J(t)$ is time dependent. Its value is random and drawn from a Brownian or short-time-correlated distribution $P[J(t)]$. 
    \item The dynamics does not preserve any ordinary conserved quantity, such as energy or particle number.
\end{enumerate}
Importantly, our theory does not require specifying the detailed form of the microscopic couplings or the Hamiltonian.

The basic idea is as follows. In the field-theoretic formulation, the OTOC involves alternating forward and backward time evolution, which is naturally represented as a path integral on four Keldysh contours [Fig.~\ref{fig:contour}(a)]. Our goal is therefore to construct an effective action $S[n,\phi]$ for the relevant degrees of freedom $(n,\phi)$ on this four-fold contour, from which the OTOC can be evaluated. The identification of these degrees of freedom is guided by an emergent strong U(1) symmetry in the noninteracting, quadratic-fermion limit, which is explicitly broken once interactions are introduced. To illustrate this point, consider dynamics generated by a time-dependent Hamiltonian quadratic in Majorana operators. The precise form of the Hamiltonian is unimportant, and in general it need not possess any ordinary conserved quantity. Nevertheless, a strong U(1) symmetry emerges after the four Keldysh contours are reinterpreted as a standard two-segment contour in a doubled Hilbert space [Fig.~\ref{fig:contour}(b)]. In this doubled description, the Hamiltonian acting on the two copies of the Hilbert space is invariant under a global rotation that mixes the Majorana fields between the two copies. The spontaneous breaking of this strong U(1) symmetry down to a weak symmetry gives rise to a slow phase mode $\phi$ associated with the strong-charge creation operator. We therefore identify this phase field $\phi$, together with its conjugate density $n$, as the relevant degrees of freedom in the effective action $S[n,\phi]$. 
More crucially, we show that the operator $n$ has the physical meaning of operator size and is directly related to the OTOC.
Intuitively, a quadratic fermionic Hamiltonian cannot change the size of a Heisenberg operator. 
This conservation of operator size gives rise to an emergent hydrodynamics governed by the strong-to-weak spontaneous symmetry breaking.

We proceed by constructing the most general effective action $S[n,\phi]$ for the Goldstone mode consistent with the general symmetry constraints of the Keldysh path integral, including normalization, unitarity, and convergence. At leading order in $\phi$,\footnote{In standard Keldysh terminology, the strong phase field $\phi$ is identified with the quantum, or response, field. The effective action can therefore be organized as a systematic expansion in powers of $\phi$.} the effective action
\begin{equation}
    \frac{S[n,\phi]}{N}=\int n\big(\partial_t+D\nabla_x^2\big)\phi+\cdots
\end{equation}
predicts diffusive spreading of the density $n$ with diffusion constant $D$, consistent with known results for operator-size dynamics in random free-fermion systems.

Now, quantum chaos requires adding $q\geq 4$-body interactions. Such terms explicitly break the emergent U(1) symmetry and introduce a mass term at lowest order for the would-be Goldstone mode. Remarkably, the form of higher-order symmetry breaking terms is tightly constrained by a new symmetry that we uncover in this work. We show that the effective action must be invariant under a combined transformation of time reversal and contour permutation. This leads to a duality condition relating $n$ and $\phi$, which must be imposed order by order in $S[n,\phi]$. This duality condition, together with the exact invariant fixed points of operator dynamics at $n=0$ (the identity operator) and $n=1$ (the global fermion parity operator), completely fixes the effective action up to order $\phi^2$:
\begin{equation}
    \frac{S[n,\phi]}{N}=\int  n\big(\partial_t+D\nabla_x^2\big)\phi+\lambda\phi n(1-n)(1-2n)+i\lambda \phi^2n(1-n)(1-2n+2n^2)+O(\phi^3).
    \label{eq:intro_S}
\end{equation}
The resulting equation of motion for the operator-size density is a noisy FKPP-type equation,
\begin{equation}
    \partial_t n-D\nabla_x^2 n=\lambda n(1-n)(1-2n)+\sqrt{\frac{2\lambda}{N}n(1-n)(1-2n+2n^2)}\,\xi(x,t),
\end{equation}
where $\xi(x,t)$ is Gaussian white noise. Interestingly, the duality relation ties the noise strength in Eq.~\eqref{eq:intro_S} directly to the Lyapunov exponent $\lambda$. Consequently, convergence of the path integral enforces the positivity of $\lambda$.

The main novelty of our work is not merely that the OTOC obeys an FKPP-type equation. Related equations have appeared previously in random circuits, Brownian Hamiltonians, and large-$N$ kinetic descriptions. Rather, our result is a symmetry-based derivation of the effective theory itself. The fermionic FKPP structure, nonlinear saturation, multiplicative noise, and the relation between the noise strength and the Lyapunov exponent all follow from the Schwinger-Keldysh consistency conditions, the structure of operator space, and the emergent duality associated with time reversal and contour permutation.

We test this general framework in a Brownian SYK chain with both quadratic and quartic interactions. A direct saddle-point expansion of the microscopic Brownian SYK model precisely reproduces the structure of the effective action~\eqref{eq:intro_S}, with the parameters $D$ and $\lambda$ determined by the microscopic couplings of the model. This provides a nontrivial check that the effective theory captures the universal dynamics of operator scrambling.

The rest of the paper is organized as follows. In Sec.~\ref{sec:U(1)}, we review the operator-to-state mapping and relate fermionic OTOCs to operator-size correlators. In Sec.~\ref{sec:eft}, we identify the strong and weak symmetry structures on the Keldysh contour and construct the corresponding effective action. In particular, we derive the emergent duality and use it to fix the minimal nonlinear action. In Sec.~\ref{sec:dynamics of OTOC}, we extract the OTOC from the effective theory and derive the noisy fermionic FKPP equation. In Sec.~\ref{sec:SYK}, we verify the effective theory in a Brownian SYK chain. We conclude with a discussion of extensions and open questions.

\section{Operator size and OTOC on Keldysh contours}
\label{sec:U(1)}

\subsection{Operator to state mapping}\label{subsec:operator_state}

Under unitary time evolution, an initially local operator $\mathcal{O}$ evolves in the Heisenberg representation as $\mathcal{O}(t)=U^\dagger(t) \mathcal{O}(0) U(t)$ into a superposition of increasingly complex operators with growing support. 
Understanding this process of operator scrambling provides a useful lens to characterize the universal properties of the underlying dynamics and quantum chaos. It is often convenient to expand the operator $\mathcal{O}(t)$ in an orthonormal operator basis $\{\Gamma_X\}$ so that $\mathcal{O}(t) = \sum_X a_X(t) \Gamma_X$. For fermionic systems, a natural choice of such an operator basis is an ordered product of Majorana fermions:
\begin{equation}\label{eq:Gamma_X}
    \Gamma_{X}\equiv \Gamma_{x_1,j_1; x_2,j_2; \ldots x_m,j_m}=\big(2i^{m-1}\big)^{m/2}\psi_{x_1,j_1}\psi_{x_2,j_2}...\psi_{x_m,j_m}\, ,
\end{equation}
where $X$ is shorthand for $\{x_1,j_1;\ldots;x_m,j_m\}$ and labels both sites and fermion species. The prefactors are chosen according to the convention $\{\psi_{x,j},\psi_{y,k}\}=\delta_{xy}\delta_{jk}$ so that each $\Gamma_X$ is Hermitian and satisfies ${\rm Tr}(\Gamma_X^\dagger \Gamma_Y)=\delta_{XY}\D$, where $\D$ is the total Hilbert-space dimension. By construction, $\Gamma_X$ in Eq.~\eqref{eq:Gamma_X} has operator size $m$. The average operator size of $\mathcal{O}(t)$ is given by
\begin{equation}
    \mathcal{N}[\mathcal{O}(t)]= \sum_X |a_X(t)|^2 \mathcal{N}[\Gamma_X]\,.
\end{equation}

It turns out that the operator size $\mathcal{N}[\mathcal{O}(t)]$ admits a compact representation using the isomorphism between operators and states in a doubled Hilbert space, as originally introduced in Refs.~\cite{jamiolkowski1972linear,choi1975completely,qi2019quantum,gu2017spread}. We start by considering two copies of the original Hilbert space, denoted by $L$ and $R$: $\H^2\equiv \H_L\otimes\H_R$. Majorana operators acting on the two copies are labeled by $\psi_{x,j}^L$ and $\psi_{x,j}^R$, respectively. The isomorphism proceeds by constructing a maximally entangled state $|0\rangle$ between $L$ and $R$. An operator $\mathcal{O}$ acting on the original Hilbert space is then mapped to a state in the doubled Hilbert space via
\begin{equation}
    \mathcal{O} \; \mapsto \; |\mathcal{O}\rangle \equiv \big(\mathcal{O}^L \otimes \mathbb{I}^R\big) \ |0\rangle\,.
\end{equation}
It is straightforward to check that the inner product in the doubled Hilbert space reproduces the inner product in the original operator space: $\langle \O_A|\O_B\rangle={\rm Tr}(\O_A^\dagger \O_B)$.

\begin{figure}[t]
    \centering
    \includegraphics[width=0.95\linewidth]{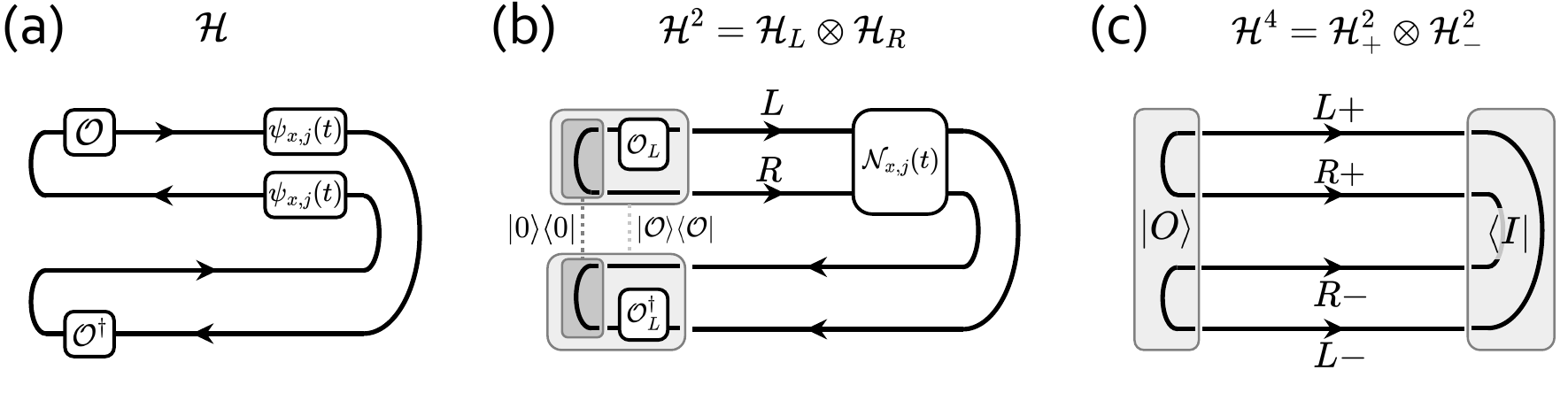}
    \caption{ (a) To compute OTOCs such as~\eqref{eq:C_1} and \eqref{eq:C_l}, the path integral must be taken along a four-fold Keldysh contour. (b) In the doubled Hilbert space $\mathcal H^2$, this four-fold contour is reduced to a standard two-fold contour. The OTOC is equivalent to the expectation value of the occupation-number operator $\mathcal N_{x,j}(t)$ in the state $|\mathcal O\rangle\equiv (\O_L\otimes\mathbb{I}_R)|0\ket$. (c) The system can be further mapped to a single directed path in the quadrupled Hilbert space $\mathcal H^4$, where the initial state $|O\rangle$ originates from the physical infinite-temperature initial state, while the final state $\langle I|$ corresponds to the trace operation in $\mathcal H^2$.}
    \label{fig:contour}
\end{figure}

The definition of $|0\rangle$ is not unique, since one can always perform local unitary transformations acting separately on the $L$ and $R$ Hilbert spaces. For concreteness, we choose the following convention for our Majorana chain model, which is diagonal in the labels~\cite{gu2017spread,maldacena2018eternal,qi2019quantum}:
\begin{equation}\label{eq:psipsi|0>}
    \big(\psi^{L}_{x,j}+i\psi^{R}_{x,j}\big)|0\ket_{x,j}=0,\qquad |0\ket\equiv\bigotimes_{x,j}|0\ket_{x,j}\,,
\end{equation}
where $x=1,2,\ldots,M$ labels the site and $j=1,2,\ldots,N$ denotes the fermion species at each site. In this construction, on the one hand, $|0\rangle$ is evidently the vacuum of the complex fermionic annihilation operator
\begin{equation}\label{eq:entangle}
    c_{x,j}\equiv \frac{1}{\sqrt{2}}\big(\psi^{L}_{x,j}+i\psi^{R}_{x,j}\big), \quad c_{x,j}|0\ket=0, \quad \forall \ x, j\,.
\end{equation}
On the other hand, Eq.~\eqref{eq:psipsi|0>} indeed implies that $|0\rangle$ is a maximally entangled state between the $L$ and $R$ Hilbert spaces. To see this, one may pair the Majorana operators in the $L$ and $R$ spaces separately into creation and annihilation operators within each space. The maximally entangled state constructed from the occupation-number bases of the two spaces then satisfies Eq.~\eqref{eq:psipsi|0>}. An explicit construction can be found in Ref.~\cite{gu2017spread}. In what follows, $c_{x,j}$ defined in Eq.~\eqref{eq:entangle}, together with its Hermitian conjugate, will be referred to as the entangled basis.

The advantage of the basis choice in Eq.~\eqref{eq:psipsi|0>} is that each basis operator $\Gamma_X$ is directly mapped to an occupation configuration of complex fermions in the entangled basis, since
\begin{equation}
    |\Gamma_X\ket\equiv\big(2i^{m-1}\big)^{m/2}\psi^L_{x_1,j_1}...\psi^L_{x_m,j_m}|0\ket=c^\hc_{x_1,j_1}...c^\hc_{x_m,j_m}|0\ket\, .
\end{equation}
In other words, each nontrivial Majorana operator in the string $\Gamma_X$ creates an excitation of the corresponding fermionic mode on top of the vacuum $|0\rangle$. Therefore, its operator size can be probed simply by the occupation-number operator for each site and species:
\begin{equation}
\N_{x,j}\equiv c_{x,j}^\hc c_{x,j}=\frac{1}{2}+i\psi^L_{x,j}\psi^{R}_{x,j} \,.
\label{eq:N_xj}
\end{equation}
For a generic operator $\O$, the expectation value of the number operator, $\langle \O|\N_{x,j}|\O\rangle=\sum_X|\a_X|^2\delta_{(x,j)\in X}$, evaluates the probability that $\O$ contains the operator $\psi_{x,j}$. Hence, the expectation value of the spatially local number operator $\N_x=\sum_j\N_{x,j}$ measures the total operator weight of $\O$ on site $x$. 

For an operator $\mathcal{O}(t)$ supported on $m$ Majorana fermions, a straightforward calculation relates the operator size to the OTOC between $\O$ and $\psi_{x,j}$ as~\cite{qi2019quantum}
\begin{equation}\label{eq:<O|N|O>}
    \bra \O(t)|\N_x|\O(t)\ket=\bra \O|\N_x(t)|\O\ket=\frac{1}{2}\sum_j \tr\left[\big|\{\psi_{x,j}(t),\O\}_m\big|^2\right],
\end{equation}
where $\{\cdots\}_m$ denotes a commutator for even $m$ and an anticommutator for odd $m$. As an example, the OTOC between a single Majorana operator initially located at the origin, $\psi_{0,k}$, and a single Majorana operator $\psi_{x,j}$ located at site $x$, averaged over the fermion species $k$ and $j$, is
\begin{equation}\label{eq:C_1}
    \mathcal{C}_{1}(x,t)=\frac{1}{N^2}\sum_{j,k} \tr\left[\big|\{\psi_{x,j}(t),\psi_{0,k}\}\big|^2\right].
\end{equation}
More generally, the OTOC between $m$ Majorana operators initially located at the origin and a single Majorana operator $\psi_{x,j}$ located at site $x$ is defined as
\begin{equation}\label{eq:C_l}
    \mathcal{C}_{m}(x,t)=
    \frac{1}{N^{m+1}}\sum_{j,k_1...k_m} \tr\left[\big|\{\psi_{x,j}(t),\psi_{0,k_1}...\psi_{0,k_m}\}_m\big|^2\right],
\end{equation}
which is the OTOC we focus on in the rest of this paper.

\subsection{Dynamics in the doubled Hilbert space}
\label{sec:double}

The evaluation of the OTOC in Eq.~\eqref{eq:<O|N|O>} can be formulated as a path integral on a Keldysh contour, which admits a simple pictorial representation [see Fig.~\ref{fig:contour}(a)]. As depicted in Fig.~\ref{fig:contour}(a), the path-integral representation of the OTOC involves two forward and two backward time contours, with operator insertions arranged in an out-of-time-ordered sequence. Using the doubled Hilbert space formalism, one can equivalently interpret the four Keldysh contours as two Keldysh contours in a doubled Hilbert space~\cite{cheng2025hydrodynamic}. More precisely, define a Hamiltonian acting on the doubled Hilbert space
\begin{equation}
    H_D\equiv H_L - H_R, \quad {\rm with} \ H_L = H \otimes \mathbb{I}, \ H_R = \mathbb{I}\otimes H^T,
    \label{eq:Hd}
\end{equation}
where the subscript $D$ stands for double. The minus sign in front of $H_R$ comes from reversing the time arrow on the second segment of the contour. By construction, we have $(H_L-H_R)|0\ket=0$ as a boundary condition. The transpose $H^T$, however, is basis dependent. With the definition of $|0\ket$ in Eq.~\eqref{eq:entangle}, one can readily verify that for an SYK-like Hamiltonian,
\begin{equation}
    H(t)=\sum_{\{q\}} H_q(t),\quad H_q(t)\equiv \frac{i^{q/2}}{q!}\sum_{\{x_i\},\{j_i\}}J_{x_1,...,x_q;j_1,...,j_q}(t)\psi_{x_1,j_1}\psi_{x_2,j_2}...\psi_{x_q,j_q}\,,
\end{equation}
the component describing the $q$-body interaction satisfies
\begin{equation}\label{eq:transpose}
    H_q^{T}= (-1)^{q/2} H_q\,.
\end{equation}
In particular, for quadratic fermion Hamiltonians with $q=2$, we have $H_2^T=-H_2$, whereas for quartic interactions with $q=4$, we have $H_4^T=H_4$.

The path integral can now be viewed as representing unitary time evolution under $H_D$ starting from the initial state $|0\rangle \langle 0|$:
\begin{equation}
    \rho(t) = U_D(t) \ |0\rangle \langle 0|\ U_D^\dagger(t)\,,
    \label{eq:rho}
\end{equation}
where $|0\rangle$ is the vacuum state in the entangled $L/R$ basis defined in Eq.~\eqref{eq:entangle} [see Fig.~\ref{fig:contour}(b)]. The advantage of this doubled Hilbert space formulation is that the OTOC can be represented as a two-point correlation function and hence evaluated using a standard Schwinger-Keldysh effective action~\cite{cheng2025hydrodynamic}.

\subsection{Emergent strong U(1) symmetry in doubled Hilbert space}\label{sec:symm&SSB}

For a generic Brownian SYK-type Hamiltonian with $q$-body interactions, there is no conserved quantity associated with a continuous symmetry, since neither particle number nor energy is conserved. However, for the special case of $q=2$, we will show that the Hamiltonian $H_D$ acting on the doubled Hilbert space has an emergent U(1) symmetry associated with global rotations that mix the fermionic operators in the $L$ and $R$ spaces.
To see this, recall that for $q=2$, we have
\begin{equation}
    H_D = H_{2,L} + H_{2,R}=\frac{i}{2}\sum_{xy,jk}J_{xy,jk}(t)\big(\psi_{x,j}^L\psi_{y,k}^L+\psi_{x,j}^R\psi_{y,k}^R\big),
\end{equation}
where, with a slight abuse of notation, we have left the tensor product with the identity operator implicit. Notice that the sign flip compared to Eq.~\eqref{eq:Hd} is due to the additional minus sign from $H_2^T=-H_2$. Now consider the U(1) transformation generated by the total occupation number in the entangled $L/R$ basis, $\mathcal{N}=\sum_{x,j}\mathcal{N}_{x,j}$ [Eq.~\eqref{eq:N_xj}],
\begin{equation}
    \U(\theta)= e^{i\theta\N}\,.
\end{equation}
Because $\mathcal{N}$ is quadratic in the Majorana fermion operators, $\U(\theta)$ acts as an SO(2) rotation on the doublet of $\psi^L_{x,j}$ and $\psi^R_{x,j}$ for each flavor at every site:
\begin{equation}\label{eq:UPsiU}
    \U(\theta) \Psi_{x,j}\,\U^\hc(\theta)=e^{i\theta \sigma^y}\Psi_{x,j},\quad \Psi_{x,j}=\binom{\psi^L}{\psi^R}_{x,j}.
\end{equation}
Since $H_D$ for $q=2$ is written as an inner product between the $\Psi_{x,j}$ fields, it follows that $H_D$ is invariant under $\U(\theta)$. In addition, since the initial density matrix $|0\rangle \langle0|$ is the vacuum state of $\mathcal{N}$, the time-evolved density matrix $\rho(t)$ is also symmetric under $\mathcal{U}(\theta)$ acting on either the forward ($+$) or backward ($-$) contour:
\begin{equation}
    \mathcal{U}(\theta)\ \rho(t) = e^{i\alpha} \rho(t), \quad \rho(t)\ \mathcal{U}(\theta)^\dagger = e^{-i\alpha} \rho(t).
\end{equation}
In terms of the symmetry of mixed states, $\rho(t)$ has an emergent strong U(1) symmetry in the doubled Hilbert space.

We make a few remarks about this emergent symmetry. First, it is a special feature of $q=2$ and is explicitly broken once interactions with $q\geq 4$ are included. Nevertheless, we use the $q=2$ limit as an anchoring point for constructing a hydrodynamic effective field theory for the associated Goldstone mode. 
In this noninteracting limit, the Goldstone mode arises from a {\it spontaneous} strong-to-weak symmetry breaking of this U(1) symmetry, which describes hydrodynamic transport of the U(1) charge injected by the initial operator in the OTOC path integral [Fig.~\ref{fig:contour}(a)]~\cite{hauser2026strong}. As will be discussed in Sec.~\ref{sec:eft} and~\ref{sec:dynamics of OTOC}, the strong charge density precisely corresponds to the local operator size, from which the behavior of OTOC can be obtained. Second, interactions with $q \geq 4$ {\it explicitly} break this emergent U(1) symmetry, which, at lowest order, leads to a mass term for the would-be Goldstone mode. 
By adding symmetry breaking terms systematically, order by order and subject to the symmetry constraints of the Keldysh path integral, we obtain an effective field theory which encodes a universal theory for the OTOC dynamics.
Finally, this strong U(1) symmetry is in fact part of a larger $O(4)$ symmetry on the four Keldysh contours~\cite{zhang2021emergent} (see next section). However, since the initial state $|0\rangle \langle0|$ is not invariant under the full O(4) symmetry, the density matrix relevant for our setup is invariant only under a smaller symmetry group.

\section{Construction of the effective field theory from strong-to-weak symmetry breaking}\label{sec:eft}

In this section, we construct an effective field theory (EFT) for operator spreading by writing down the most general effective action on the Schwinger-Keldysh contour. We begin by identifying the generators of the strong and weak symmetries in terms of the fields living on the four-fold Schwinger-Keldysh contour, which constitute the degrees of freedom of the EFT. We then list the general constraints imposed on the effective action by unitarity and related consistency conditions. In the absence of explicit U(1) symmetry breaking induced by $q\geq 4$ interactions, these constraints lead to a hydrodynamic theory for the Goldstone mode associated with spontaneous breaking of the strong symmetry. We next incorporate $q\geq 4$ interactions, which {\it explicitly} break the U(1) symmetry and therefore generate a mass term for the would-be Goldstone mode. This mass term plays the role of Lyapunov exponent, converting diffusive operator spreading into chaotic growth. Remarkably, the structure of higher-order symmetry breaking terms are further constrained by a combination of time reversal and contour permutation symmetries. As we show, these symmetries imply a duality condition relating the fields in the effective action. Combining these ingredients, we arrive at the general form of the effective action.

\subsection{Degrees of freedom of the EFT}

As discussed in Sec.~\ref{sec:double}, the four-fold Keldysh contour on which the effective action for the OTOC lives can be interpreted as a regular two-fold Keldysh contour in a doubled Hilbert space, with two forward ($+$) and two backward ($-$) contours evolving under $H_D$. At $t=t_f$, the $+$ and $-$ fields are related by boundary conditions. Similarly to Eq.~\eqref{eq:psipsi|0>}, we can define a state $|I\rangle$ at the $t=t_f$ boundary, which is a maximally entangled state between the two $+$ and two $-$ contours. The boundary condition that determines the state $|I\rangle$ is chosen as
\begin{equation}\label{eq:psiLpsiR|I>=0}
    \big(\psi^{L+}_{x,j}-i\psi^{L-}_{x,j}\big)|I\ket=0,\quad \big(\psi^{R+}_{x,j}+i\psi^{R-}_{x,j}\big)|I\ket=0.
\end{equation}
This boundary condition implies $\N^+|I\rangle=\N^-|I\rangle$, which is the contour representation of taking the trace at $t_f$. Written explicitly in the occupation-number basis of the $L/R$ entangled basis defined in Eq.~\eqref{eq:entangle}, we find
\begin{equation}\label{eq:|I>=|00>+|11>}
    |I\ket=\bigotimes_{x,j}\big(|00\ket_{x,j}+i|11\ket_{x,j}\big),\quad |00\ket=|0\ket^+|0\ket^-,\quad |11\ket=c^{+\hc}c^{-\hc}|00\ket.
\end{equation}
The state is normalized as $\bra I|I\ket=2^{MN}$. With the final state $|I\rangle$ specified, the Schwinger-Keldysh evolution can be unfolded into a transition amplitude in the quadrupled Hilbert space, $\H^4=\H_+^2\otimes \H_-^2$. Consequently, the initial density matrix is vectorized as $|O\rangle=|0\rangle|0\rangle$, and the two-fold Keldysh contour in Fig.~\ref{fig:contour}(b) is mapped to the single directed path from the initial state $|O\rangle$ to the final state $\langle I|$ shown in Fig.~\ref{fig:contour}(c).

\begin{figure}[t]
    \centering
    \includegraphics[width=0.9\linewidth]{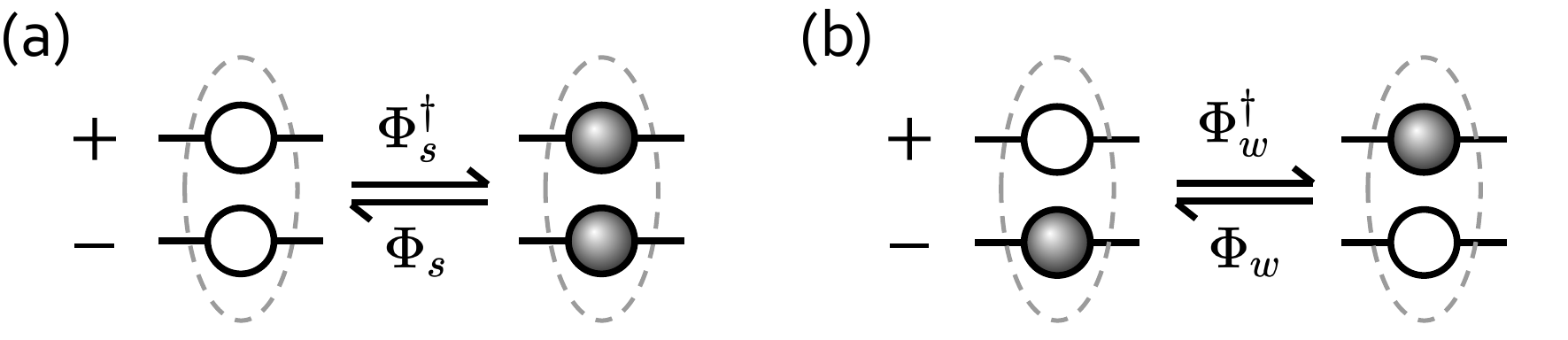}
    \caption{The physical interpretations of $\Phi$-operators defined in \eqref{eq:Phi_sxPhi_wx}. (a) $\Phi_s$ ($\Phi_s^\hc$) annihilates (creates) fermions in pairs on the $+$ and $-$ contours. (b) $\Phi_w$ ($\Phi_w^\hc$) moves a fermion from the $+$ ($-$) contour to the $-$ ($+$) contour. Fermion species and site indices are suppressed.}
    \label{fig:PhiPhihc}
\end{figure}

In this way, the density matrix $\rho(t)$ in Eq.~\eqref{eq:rho} can be vectorized as a state $|\rho\rangle$ living in the quadrupled Hilbert space. A pair of symmetry rotations acting on the bra and ket of the density matrix is then represented as
\begin{equation}\label{eq:rho_symmetry_vectorized}
    \U(\theta_+)\,\rho\,\U^\hc(\theta_-)
    \quad\mapsto\quad
    e^{i\theta_+\N^+-i\theta_-\N^-}|\rho\ket .
\end{equation}
Motivated by this representation of the symmetry transformation, we further combine the two U(1) generators on the $+$ and $-$ contours into generators of the strong and weak symmetries,
\begin{equation}\label{eq:strong,weak_DHS}
    \N_s=\frac{\N^++\N^-}{2},\quad \N_w=\N^+-\N^-.
\end{equation}
The weak generator $\N_w$ implements the usual conjugation of the density matrix, corresponding to $\theta_+=\theta_-$. By contrast, $\N_s$ generates the relative rotation of the two sides of the density matrix and is the contour representation of the strong symmetry. One can define ladder operators that change the strong and weak symmetry charges on each site:
\begin{equation}\label{eq:Phi_sxPhi_wx}
    \Phi_{s,x}\equiv \sum_jc_{x,j}^-c^+_{x,j}\,, \qquad \Phi_{w,x}\equiv \sum_jc_{x,j}^{-\hc}c_{x,j}^+\,,
\end{equation}
where the sum runs over fermion species on each site. One can readily check that these operators satisfy an $su(2)\oplus su(2)$ algebra,
\begin{equation}\label{eq:su2su2}
    \big[\N_{\sigma,x},\Phi_{\sigma',y}\big]=-C_{\sigma}\Phi_{\sigma,x}\delta_{\sigma\sigma'}\delta_{xy}\,,\quad \quad \big[\Phi_{\sigma,x}^\hc,\Phi_{\sigma',y}\big]=\big(2C^{-1}_{\sigma}\N_{\sigma,x}-N\delta_{\sigma s}\big)\delta_{\sigma\sigma'}\delta_{xy}\,,
\end{equation}
where $\sigma$ and $\sigma'$ label the strong $(s)$ and weak $(w)$ sectors, with $C_{s}=1$ and $C_w=2$. The physical interpretation of these ladder operators is clear and is illustrated in Fig.~\ref{fig:PhiPhihc}. The operator $\Phi_s\;(\Phi_s^\hc)$ annihilates (creates) a fermion simultaneously on the $+$ and $-$ contours, thus changing $\mathcal{N}_s$ while keeping $\mathcal{N}_w$ unchanged. In contrast, $\Phi_w\;(\Phi_w^\hc)$ transfers a fermion from the $+(-)$ contour to the $-(+)$ contour, changing $\mathcal{N}_w$ while keeping $\mathcal{N}_s$ fixed. The strong and weak operators therefore commute with each other. To take the continuum limit of the fields corresponding to these operators, we normalize them by $N$ in the following discussion:
\begin{equation}
    n_{\sigma,x}\equiv\frac{\N_{\sigma,x}}{N},\quad
    \varphi_{\sigma,x}\equiv\frac{\Phi_{\sigma,x}}{N},\quad
    \varphi^\hc_{\sigma,x}\equiv\frac{\Phi^\hc_{\sigma,x}}{N}.
    \label{eq:nvarphi}
\end{equation}

Vectorizing the density matrix, the von Neumann equation becomes 
$i\partial_t |\rho\ket = \big(H_{D,+}(t) - H_{D,-}(t)\big) |\rho\ket$, where $H_{D,\pm}(t)$ denotes the left (right) action of $H_D$ on the density matrix. 
For instance, they are given explicitly as $H_{D,+} = H_D$ and $H_{D,-} = H_D^T$ for the SYK-like Hamiltonian in Eq.~\eqref{eq:transpose}.  
The effective theory should be understood as a collective-field representation of the transition amplitude on the quadrupled Hilbert space. In the absence of operator insertions, the generating functional takes the form
\begin{eqnarray}
    \overline{Z} = \overline{\big< I \big|U_Q(t_f,t_i)  \big| O \big> }\,, \qquad U_Q(t_f,t_i) \equiv \mathscr T \exp \Big[-i \int_{t_i}^{t_f} ds\, \big(H_{D,+}(s) - H_{D,-}(s)\big) \Big] \,,
\end{eqnarray}
where the overline denotes the ensemble average of random couplings in the microscopic Hamiltonian, and the subscript $Q$ stands for quadruple.
The transition amplitude can equivalently be written as a fermionic coherent-state path integral, 
\begin{eqnarray}
    Z = \int \D\Psi e^{i S_{\rm mic}[\Psi]} \,,
\end{eqnarray}
where $\Psi$ denotes the fermionic fields on the four contours. 
We now introduce collective fields $Q_\a\equiv(\varphi_{\sigma,x}, \varphi_{\sigma,x}^\dagger, n_{\sigma,x})$ associated with the fermion bilinear $B_\a \equiv (\Phi_{\sigma,x},\Phi_{\sigma,x}^\dagger,\mathcal N_{\sigma,x})$ and impose them by Lagrange multipliers $\Sigma^\a$, 
\begin{eqnarray}
    1 = \int \D Q \D \Sigma \exp \Big[ i N  \Sigma^{\a} \Big(Q_{\a} - \frac{B_{\a}}N  \Big)  \Big]  \,,
\end{eqnarray}
where the sum over $\a$ is suppressed. We can then integrate out the microscopic fermions to arrive formally at
\begin{eqnarray}
    e^{iS_{\rm eff}[Q]}  \equiv  \int \D \Sigma e^{iN  \Sigma^\a Q_\a +i W[\Sigma]} \,, \quad e^{iW[\Sigma]} \equiv \int \D\Psi \overline{e^{iS_{\rm mic}[\Psi] - i  \Sigma^\a B_\a}} \,. 
\end{eqnarray}
To allow an effective action approach in the following, we assume the microscopic fermion action admits a large $N$ structure, $W[\Sigma] \propto N $, so that the collective-field path integral is governed by a saddle point. 
Rather than deriving the resulting low-energy action from the detailed microscopic Hamiltonian, we construct it from the fluctuations around the collective saddle and the general consistency conditions of the Schwinger-Keldysh path integral. 

The reference saddle relevant for the OTOC generating functional is fixed kinematically by the initial and final states. Since the contour Hamiltonian acts trivially on the initial and final states, the time evolution satisfies
\begin{eqnarray}
    U_Q |O\ket = |O\ket \,, \qquad \bra I | U_Q = \bra I | \,.
\end{eqnarray}
It follows that the source-free saddle-point value of an operator $\mathcal O$ is simply the normalized transition matrix element,
\begin{eqnarray} \label{eq:bracket}
    \bra \mathcal O(t) \ket \equiv \frac{\bra I | U_Q(t_f, t) \mathcal O U_Q(t, t_i) | O \ket}{\bra I | U_Q(t_f, t_i)  | O \ket} = \frac{\bra I | \mathcal O  | O \ket}{\bra I |  O \ket} \,.
\end{eqnarray} 
Hence, the saddle-point values of the operators in \eqref{eq:nvarphi} that are pertinent to the OTOC generating functional are
\begin{equation}\label{eq:<Phi>=-iN}
    \bra \varphi_{s,x}^\hc\ket=-i,\qquad
    \bra\varphi_{s,x}\ket=\bra n_{s,x}\ket=0,\qquad
    \bra \varphi_{w,x}^\hc\ket=\bra\varphi_{w,x}\ket=\bra n_{w,x}\ket=0.
\end{equation}
It is worth noting that the brackets in Eq.~\eqref{eq:bracket} denote transition symbols rather than expectation values in a positive density matrix. The initial ket and final bra are distinct boundary states and are not Hermitian conjugates of one another. Consequently, the classical symbols of $\varphi$ and $\varphi^\dagger$ are independent complex variables.

The strong-sector variables should be understood as coordinates on the low-energy spin orbit selected by the saddle. The three operators $\Phi_s^\hc$, $\Phi_s$, and $\N_s$ form an $su(2)$ algebra. At the leading large-$N$ saddle relevant for the OTOC generating functional, the strong sector is projected to a nonzero fixed-Casimir orbit with $\bra J_s^2\ket/N^2=1/4$. Therefore the corresponding classical phase space is two-dimensional, and we choose a Darboux pair $n_s,\phi_s$ to parametrize it as
\begin{equation}\label{eq:delta_n_delta_phi}
        n_s=n_s ,\qquad 
        \vphi_s=in_se^{-i\phi_s}, \qquad 
        \vphi_s^\hc=-i(1-n_s)e^{i\phi_s}\,,
\end{equation}
where the $n_s$ dependence is determined by the saddle-point values and the $su(2)$ algebra. A more explicit derivation of these classical symbols is given in Appendix~\ref{app:nphi_parameterization}. Indeed, this parametrization obeys the leading large-$N$ fixed-Casimir relation
\begin{equation}
    \vphi_s^\hc\vphi_s+\Big(n_s-\frac{1}{2}\Big)^2=\frac{1}{4}.
\end{equation}
Thus the three normalized collective variables are not treated as independent fields; they are the classical symbols of the spin generators restricted to the saddle orbit.

The weak sector is expanded around a different saddle, where the one-point functions of $\vphi_w$, $\vphi_w^\hc$, and $n_w$ vanish. Since this saddle has zero Casimir, it should not be viewed as an ordinary fixed $j=0$ spin coherent-state orbit, which would be trivial. Rather, the weak variables parametrize the low-energy complexified zero-Casimir fluctuation around that saddle, with $n_w$ playing the role of the amplitude variable:
\begin{equation}\label{eq:delta_n_delta_phi_weak}
    n_w=n_w ,\qquad
\vphi_w=\frac{i}{2}n_we^{-2i\phi_w}, \qquad
\vphi_w^\hc=\frac{i}{2}n_we^{2i\phi_w}\,.
\end{equation}
The corresponding weak parametrization has vanishing normalized classical Casimir, as discussed in the Appendix~\ref{app:nphi_parameterization}.

Therefore, $n_s$ and $n_w$ are the charge densities conjugate to the phase fields, although their geometric roles are different: $n_s$ is a coordinate on the nonzero fixed-Casimir orbit, while $n_w$ is the amplitude emerging from the zero-Casimir saddle. The two canonical commutation relations are
\begin{equation}\label{eq:[phi,n]}
    [\phi_{s,x},n_{s,y}]=\frac{i}{N}\delta_{xy},
    \qquad
    [\phi_{w,x},n_{w,y}]=\frac{i}{N}\delta_{xy}.
\end{equation}
These relations show explicitly that $1/N$ plays the role of an effective $\hbar$, so the effective action is proportional to $N$ and the saddle-point expansion is controlled in the large-$N$ model.

Although the final EFT will be formulated primarily in the Keldysh basis (equivalently, the $s$ and $w$ basis), the constraints on the effective action are more transparent in the \(+/-\) basis. Using the convention in Eq.~(\ref{eq:strong,weak_DHS}), the inverse relation is $\N^\pm=\N_s\pm\N_w/2$. Correspondingly, the phase fields on the two contours are $\phi_\pm=\phi_w\pm\phi_s/2$. This choice is consistent with the canonical one-form on the Schwinger-Keldysh contour,
\begin{equation}
    \N^+\partial_t\phi_+-\N^-\partial_t\phi_-=\N_s\partial_t\phi_s+\N_w\partial_t\phi_w.
\end{equation}
The right-hand side follows from the commutation relation \eqref{eq:[phi,n]}, whereas the minus sign on the left-hand side arises due to reversing the time arrow on the $-$ contour.

Consequently, combining $\phi_s=\phi_+-\phi_-$ and $\phi_w=(\phi_++\phi_-)/2$ with the commutators in Eq.~(\ref{eq:[phi,n]}), we obtain the action of the contour transformations in Eq.~(\ref{eq:rho_symmetry_vectorized}) on the strong and weak phase variables:
\begin{equation}
    \phi_s\to\phi_s+\theta_+-\theta_-,\quad
    \phi_w\to\phi_w+\frac{\theta_++\theta_-}{2}.
\end{equation}
Under the weak U(1) symmetry $\theta_+=\theta_-$, the $\phi_s$ field remains unchanged and the $\phi_w$ field transforms; under the strong symmetry, $\phi_s$ transforms accordingly. Thus, $e^{i\phi_s}$ carries strong U(1) charge but not the weak charge.

\subsection{EFT for the Goldstone mode}

Having identified the appropriate degrees of freedom as $\phi_s$ and $\phi_w$ (or equivalently $\phi_+$ and $\phi_-$), we are now in a position to construct an effective action $S[\phi_s,\phi_w]$. In the absence of additional conserved quantities, such as energy conservation, we postulate, in analogy with conventional hydrodynamics, that these fields are the only relevant slow modes of the theory. The effective action $S[\phi_s,\phi_w]$ should then be understood as the result of integrating out all fast modes while keeping $(\phi_s,\phi_w)$ fixed as background fields. Moreover, as we will show in Sec.~\ref{sec:dynamics of OTOC}, these are precisely the degrees of freedom that control operator spreading and the dynamics of the OTOC.

We list some of the general constraints that must be imposed on the effective action $S[\phi_s, \phi_w]$~\cite{liu2018lectures,crossley2017effective}:

\begin{itemize}
    \item Normalization condition: when taking $\phi_+=\phi_-=\phi$, the partition function on the Keldysh contour satisfies:
    \begin{equation}
        e^{iS[\phi_+=\phi_-=\phi]}={\Tr}\big[U_{\rm fast}(\phi)\ \rho_0 \ U_{\rm fast}(\phi)^\dagger\big]=1\,,
    \end{equation}
    where the slow modes $(\phi_+,\phi_-)$ are treated as a fixed background or external source for the fast modes, and $U_{\rm fast}(\phi)$ denotes the evolution operator of the fast modes in the background of the slow fields $\phi$. This implies
    \begin{equation}
    S[\phi_+=\phi_-=\phi]=0\,, \quad  {\rm or} \quad S[\phi_s=0, \phi_w]=0\,.
    \label{eq:normalization}
    \end{equation}
    This implies that each term in the effective action must contain at least one factor of the strong field $\phi_s$.
    
    \item Reflection symmetry under complex conjugation: taking the complex conjugation of the partition function effectively exchanges the forward and backward contours, which leads to the following reflection symmetry on the effective action:
    \begin{equation}
        S[\phi_+,\phi_-]^*=-S[\phi_-,\phi_+]\,,
        \qquad
        S[\phi_s,\phi_w]^*=-S[-\phi_s,\phi_w]\,.
    \label{eq:reflection}
    \end{equation}
    An important consequence of this reflection symmetry is that the coupling in front of any term in $S[\phi_s, \phi_w]$ that is even under $\phi_s \rightarrow -\phi_s$ must be purely imaginary.
    
    \item Convergence condition: since the effective action arises from integrating out fast modes, terms with complex couplings will in general appear. In this case, convergence of the path integral requires
    \begin{equation}
    \mathrm{Im}\,S\geq 0\,.
    \label{eq:convergence}
    \end{equation}
\end{itemize}
Since the initial state is not thermal in general, we do not impose Kubo-Martin-Schwinger (KMS) condition. In Appendix \ref{app:SK_constraints} we present a derivation of the above constraints.

When the strong symmetry is spontaneously broken while the weak symmetry is preserved,
the effective action must be further invariant under~\cite{crossley2017effective, huang2025hydrodynamics} 
\begin{equation}
    \phi_s(x,t)\to\phi_s(x,t)+\alpha,\qquad
    \phi_w(x,t)\to\phi_w(x,t)+f(x).
\label{eq:ssb_constraint}
\end{equation}
That is, one is free to locally change the phase associated with the unbroken weak U(1) symmetry, but the phase associated with the broken strong U(1) symmetry can only be assigned globally. Therefore, $S[\phi_s, \phi_w]$ can only depend on $\partial_t\phi_w$, $\partial_t\phi_s$, and spatial derivatives of $\phi_s$.

We can now write down $S[\phi_s, \phi_w]$ as an expansion in powers of $\phi_s$ and its gradient, in a similar spirit as an expansion in powers of the quantum field in constructing Schwinger-Keldysh effective actions, subject to the general constraints listed above. To quadratic order in $\phi_s$, we obtain
\begin{equation}
    S[\phi_s,\phi_w]=\int \chi\partial_t\phi_w\big(\partial_t+D\nabla_x^2\big)\phi_s+i\eta(\nabla_x\phi_s)^2.
\end{equation}
In particular, convergence of the path integral requires $\eta\geq 0$. It turns out that for our purpose, it is more convenient to write down the effective action in phase space, also known as the Martin–Siggia–Rose (MSR) formulation~\cite{PhysRevA.8.423}.
In terms of the conjugate variable of $\phi_s$, we have $\N_s=\partial \L/\partial\dot\phi_s=\chi\partial_t\phi_w$. Thus, the MSR effective action takes the form
\begin{equation}\label{eq:MSR_free}
    S[\N_s,\phi_s]=\int \N_s\big(\partial_t+D\nabla_x^2\big)\phi_s+i\eta(\nabla_x\phi_s)^2\,.
\end{equation}
We recognize Eq.~(\ref{eq:MSR_free}) as the MSR action describing the diffusion of charge density $\mathcal{N}_s$, with $D$ being the diffusion constant and $\eta$ the noise. As we will explain in Sec.~\ref{sec:spreading_free}, this effective action precisely describes the OTOC in free fermion systems without translation symmetry.

\subsection{Quantum chaos as an explicit symmetry breaking}

The above construction yields an effective action where operators spread out only diffusively rather than ballistically. This is expected since free fermion systems do not exhibit quantum chaotic behavior. To derive a theory for operator spreading in genuinely chaotic quantum many-body systems, we consider turning on generic interactions with $q\geq 4$. As we have explained in Sec.~\ref{sec:symm&SSB}, such interactions explicitly break the U(1) symmetries that we started from. Therefore, a mass term is generated for the diffusive Goldstone mode. Moreover, the constraint~(\ref{eq:ssb_constraint}) must be dropped, 
which indicates that terms depending on $\phi_s$ itself are allowed in the effective action. For convenience, we suppress the $s$-index from now on. The quadratic order action is
\begin{equation}\label{eq:S/N=lambda}
    \frac{S[n,\phi]}{N}=\int n\big(\partial_t+\lambda+D\nabla_x^2\big)\phi+i\xi \phi^2+\cdots,
\end{equation}
where $\cdots$ denotes higher order terms in $\phi$ and $n$. Ignoring the noise term $i\xi\phi^2$ for the moment, the saddle point solution of this action takes the form
\begin{equation}\label{eq:light-cone,Lyapunov}
n(x,t)\sim e^{\lambda t-\frac{x^2}{4Dt}}\simeq e^{2\lambda(t-x/v_B)}\,,
\end{equation}
for an initially localized source, and the second equality is expanded near the right-moving front $x=v_Bt$. This solution describes a ballistically propagating density with a butterfly velocity $v_B=\sqrt{4\lambda D}$ and an exponential in time growth near the front~\cite{cheng2025hydrodynamic}, which is reminiscent of the universal behavior of OTOC in large-$N$ theories. As we will explain in Sec.~\ref{sec:spreading_free}, the physical interpretation of the saddle point solution $n(x,t)$ is precisely the OTOC. Remarkably, the Lyapunov exponent $\lambda$ arises in the EFT as a mass term for the would-be Goldstone mode due to an explicit symmetry breaking of the U(1) symmetry of the noninteracting theory.

In fact, the effective action captures more than the butterfly velocity and the early-time exponential growth of the operator size. By going beyond quadratic order, it can also describe the late-time saturation dynamics of the OTOC, as we now demonstrate. We proceed by promoting the constants $\lambda$ and $\xi$ in Eq.~(\ref{eq:S/N=lambda}) to density-dependent functions $\Lambda[n]$ and $\Xi[n]$. Their forms are constrained by the fixed points of the operator-size dynamics. In particular, there are two generic fixed points: $n=0$, corresponding to the trivial identity operator, and $n=1$, corresponding to the fermion parity operator. Both must be left invariant by the dynamics. Therefore, both the deterministic growth term $n\Lambda[n]$ and the noise strength $\Xi[n]$ must vanish at $n=0$ and $n=1$. The simplest choice consistent with this requirement is $n\Lambda[n]=\lambda[n]n(1-n)$ and $\Xi(n)=\xi[n] n(1-n)$, which leads to the following form of the effective action
\begin{equation}\label{eq:MSR_int}
    \frac{S[n,\phi]}{N}=\int n\big(\partial_t+D\nabla_x^2\big)\phi +\lambda[n] n(1-n)\phi+i\xi[n] n(1-n)\phi^2+\cdots.
\end{equation}
Notice that the same fixed-point condition also forbids the noise term $i\eta(\nabla\phi)^2$ in Eq.~(\ref{eq:MSR_free}), since this term does not vanish at the fixed points.

So far, the construction leading to the effective action in Eq.~(\ref{eq:MSR_int}) has been completely general. However, it remains unsatisfactory in several respects. First, can the functional forms of $\lambda(n)$ and $\xi(n)$ be further constrained, at least to the first few lowest orders in $n$? Second, operator growth requires $\lambda>0$, but this positivity is not naturally enforced by the effective action constructed so far. Third, $\lambda(n)$ and $\xi(n)$ currently appear as independent parameters. While convergence of the path integral only requires $\xi(n)>0$, it provides no further constraint on the noise strength. In the next section, we address these questions by uncovering an emergent duality relation between the $n$ and $\phi$ fields. Remarkably, this duality relates $\lambda(n)$ and $\xi(n)$, and naturally enforces the positivity of the Lyapunov exponent $\lambda$.

\subsection{Contour permutation and emergent duality condition}

In this section, we uncover another important symmetry that must be respected by the effective action. This symmetry combines time reversal with a permutation of the fields on the four Keldysh contours. At a high level, time reversal reverses the arrow of time, thereby exchanging the `initial' and `final' states $|O\rangle$ and $|I\rangle$ and changing the partition function. This change can be undone by a relabeling, or permutation, of the four contours, which exchanges $|O\rangle$ and $|I\rangle$ once again and restores the original partition function. Nevertheless, the combined operation acts nontrivially on the degrees of freedom appearing in the effective action. In particular, we show that it gives rise to an emergent duality between the $n$ and $\phi$ fields, which the effective action must obey. As we will demonstrate, this duality not only fixes the functional forms of $\lambda[n]$ and $\xi[n]$ up to order $\phi^2$, but also directly relates the noise strength to the Lyapunov exponent, and further enforces the positivity of the latter.

Recall that our effective action emerges from the partition function of some microscopic random Hamiltonian, averaged over the couplings. This partition function is represented on four Keldysh contours with two $+$ and two $-$ contours, with time evolution generated by a Hamiltonian $H_D$ in the doubled Hilbert space defined in Eq.~(\ref{eq:Hd}). One can equivalently think of this as a usual path integral representing the transition amplitude from $|O\rangle$ to $|I\rangle$, under a Hamiltonian now acting on {\it four} copies of the original Hilbert space, simply denoted as $H_Q(t)=\sum_q J_q(t)h_{Q,q}$. Using this notation, the ensemble averaged partition function can be written schematically as 
\begin{equation}
    \overline{Z} = \int \prod_j \D J_j P[J_j]\ \big< I\big|\mathcal{T}e^{-i\int \sum_q J_q(s) h_{Q,q}\,ds}\big|O\big>,
\end{equation}
where we make the time-dependent random couplings $J_q(t)$ explicit, which are drawn from a distribution $P[J_j]$.\footnote{We take $P[J_j]$ to be a Gaussian distribution of the form $P[J_j]\propto\exp\big(-\frac{1}{2\sigma_j^2}\int dtdt'J_j(t) f^{-1}_j(t-t')J_j(t')\big)$ with the inverse kernel $f_j^{-1}$ defined by $f_j(t)*f_j^{-1}(t')=\delta(t-t')$. For Brownian random models, $f_j(t-t')$ is proportional to $\delta(t-t')$. But more generally, we only assume that it is a function of $|t-t'|$.} We start by considering how $\overline{Z}$ transforms under time reversal.

\textit{Time reversal}. Under time reversal, the partition function transforms as
\begin{equation}
\begin{split}
    \T\overline{Z} \T^{-1}=\overline Z^*=\int \prod_j \D J_j P[J_j]\ \big< O\big|\mathcal{T}e^{i\int \sum_q J_q(-s) h_{Q,q}\, ds}\big|I\big>,
\end{split}
\end{equation}
where the integral ranges are $-\infty$ to $+\infty$ such that the time-ordering is recovered upon replacing $s$ with $-s$.
Consequently, $\mathcal{T}$ maps the partition function to that of a time-reversed process exchanging the initial and final states, and evolving under the Hamiltonian
\begin{equation}
    H_Q(t) \rightarrow - H_Q(-t).
\end{equation}
Apparently, the time-reversed partition function does not have the same functional form as $\overline{Z}$. Nevertheless, the original partition function can be restored upon permuting the contours, which we discuss now.

\begin{figure}[t]
    \centering
    \includegraphics[width=1\linewidth]{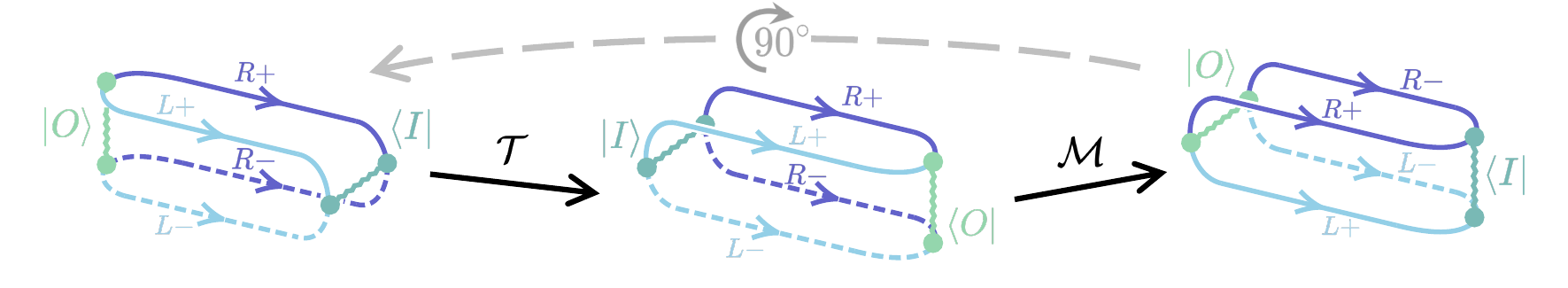}
    \caption{The contour structure is invariant under a combination of time reversal ($\T$) and a permutation ($\mathcal{M}$) of the contour labels.}
    \label{fig:MPT}
\end{figure}

\textit{Contour permutation}. Consider the following transformation on the Majorana fields associated with the four contours:
\begin{equation}
    \mathcal{M}: \quad \psi^{L-}\to\psi^{L+}\to\psi^{R+}\to\psi^{R-}\to-\psi^{L-}.
\end{equation}
As depicted in Fig.~\ref{fig:MPT}, this amounts to a relabeling or permutation of the four Keldysh contours. The transformation satisfies $\M^{-1}=\M^\hc$.
An immediate observation is that the two boundary states $|O\rangle$ and $|I\rangle$ are exchanged under this transformation, since they connect $L/R$ and $+/-$ contours respectively. Indeed, one can readily check that the boundary conditions defining the two states in Eq.~(\ref{eq:psipsi|0>}) and~(\ref{eq:psiLpsiR|I>=0}) are precisely flipped under the transformation $\mathcal{M}$. Next, it is also straightforward to check that the components of the Hamiltonian $H_Q(t)$ acting on quadrupled Hilbert space transform as 
\begin{equation}\label{eq:PHP}
    \M \ h_{Q,q}\ \M^{-1}=-(-1)^{q/2}h_{Q,q},
\end{equation}
where the factor $(-1)^{q/2}$ can be traced back to the boundary condition~(\ref{eq:psiLpsiR|I>=0}) relating the $+/-$ contours (see Appendix~\ref{app:duality} for a detailed derivation).

Hence, the combination of the time reversal transformation and the contour permutation, denoted by $\M\T$, transforms the averaged partition function as
\begin{equation}
    \begin{split}
        \M\T \ \overline{Z} \ (\M\T)^{-1}=\int \prod_j \D J_j P[J_j]\big< I\big|\mathcal{T}e^{-i\int \sum_q [(-1)^{q/2}J_q(-s)] h_{Q,q}\,ds}\big|O\big>
    \end{split}
\end{equation}
We find that the net effect of the $\mathcal{MT}$ transformation is simply the replacement $J_q(t) \rightarrow (-1)^{q/2}J_q(-t) = \pm J_q(-t)$, under which the probability distribution $P[J_j]$ is invariant. Thus, the ensemble averaged partition function is invariant under the joint action of $\mathcal{MT}$:
\begin{equation}\label{eq:Z=Z*}
    \M\T \ \overline{Z}\ (\M\T)^{-1}=\overline{Z}.
\end{equation}
Notice that the fields $n$ and $\phi$ transform nontrivially under $\mathcal{MT}$.
This immediately implies the following symmetry of the effective action:
\begin{equation}
    S[n(t),\phi(t)]=S\big[\widetilde{n}(-t),\widetilde{\phi}(-t)\big],
\end{equation}
where we denote by $\widetilde{n}(-t)$ and $\widetilde{\phi}(-t)$ the new fields under $\mathcal{MT}$ transformation. To find the transformation of $n$ and $\phi$ fields under $\mathcal{MT}$, we go back to the original fields in Eq.~(\ref{eq:nvarphi}) in terms of fermion bilinears.
As we detail in Appendix~\ref{app:duality}, the fermion bilinears transform as
\begin{equation}
    n(t)-\frac{1}{2}\xrightarrow{\,\mathcal{MT}\,}\frac{i}{2}\big(\varphi(-t)-\varphi^\hc(-t)\big),\qquad \frac{i}{2}\big(\varphi(t)-\varphi^\hc(t)\big)\xrightarrow{\,\mathcal{MT}\,}n(-t)-\frac{1}{2}\,.
\end{equation}
Then, by Eq.~(\ref{eq:delta_n_delta_phi}), we find the following duality between the $n$ and $\phi$ fields:
\begin{equation}
    n\xleftrightarrow{\; \mathcal{MT} \;} in\sin\phi-\frac{1}{2}\big(e^{i\phi}-1\big)\equiv \chi(n,\phi), \quad t\xrightarrow{\,\mathcal{MT}\,} -t\,,
    \label{eq:duality}
\end{equation}
under which the effective action must be invariant.

We begin by examining the effective action in Eq.~(\ref{eq:MSR_free}) for the Goldstone mode in the presence of the emergent U(1) symmetry. Consider the linearized form of the duality transformation in Eq.~(\ref{eq:duality}),
\begin{equation}
    n(t) \rightarrow -\frac{i}{2} \phi(-t), \quad  \phi(t) \rightarrow 2in(-t)\,.
\end{equation}
It is straightforward to verify that the effective action in Eq.~(\ref{eq:MSR_free}) is invariant under this transformation, up to boundary terms, provided that $\eta=0$. Thus the duality forbids the $(\nabla\phi)^2$ noise term. The more powerful constraint arises in the quantum chaotic case, where the U(1) symmetry is explicitly broken and a mass term is generated. As we will show, the duality condition then imposes nontrivial constraints on the allowed symmetry breaking terms.

Consider the following symmetry breaking interaction term, truncated at order $K+2$, which is manifestly invariant under the duality transformation: 
\begin{equation}
    \frac{\L_{\rm int}}{N}=i\sum_{p=1}^K\sum_{q=1}^{\min(p,2)}\mu_{pq}\left(n^p\chi^q+n^q\chi^p\right).
\end{equation}
Notice that $\chi[n,\phi]$ satisfies $\chi^*[n,\phi]=\chi[n,-\phi]$, the reflection condition $S^*[n,\phi]=-S[n,-\phi]$ thus requires the coefficients $\mu_{pq}$ to be real. Moreover, the normalization condition $S[n,\phi=0]=0$ implies that the summation over both $p$ and $q$ start from $1$. As in the standard Keldysh formulation, the effective action can be organized as an expansion in powers of the quantum field $\phi$. Terms of order $O(\phi^2)$ describe Gaussian noise, whereas terms of order $O(\phi^3)$ and higher encode higher noise cumulants and hence non-Gaussian fluctuations. Because $\chi[n,\phi]=O(\phi)$, retaining $q \leq 2$ is sufficient to capture all contributions at $O(\phi^2)$. Higher noise cumulants can be incorporated systematically by extending the expansion, but we neglect them in what follows.

We now expand the symmetry breaking interaction term in powers of $n$ and $\phi$, organizing the result order by order in $\phi$. A direct consequence is that, since the leading contribution of $\chi[n=1/2,\phi]$ is of order $O(\phi^2)$, the deterministic term associated with the term linear in $\phi$ vanishes, and therefore $n=1/2$ must be a fixed point. This is consistent with the fact that this point describes the equilibrium operator distribution with average operator weight $n=1/2$. The minimal truncation $K=1$ gives the effective action for the most basic diffusion-reaction process, whose interaction term is given by
\begin{equation}
    \frac{\L_{\rm int}}{N}=\lambda\phi n(1-2n)+\frac{i\lambda}{2}\phi^2n+O(\phi^3)\,,
\end{equation}
where $\lambda=\mu_{11}$. However, this action does not satisfy the fixed-point condition $S[n=1,\phi]=0$, and is therefore unphysical in fermionic models. The minimal truncation satisfying this condition is $K=2$. Still choosing $\mu_{11}=\lambda$, the minimal interaction term up to quadratic order in $\phi$ is given by
\begin{equation}
\begin{split}
    \frac{\L_{\rm int}}{N}= & \,\phi n\Big(\lambda+\frac{\mu_{21}}{2}n\Big)(1-2n)+\frac{i\phi^2n}{4}\Big(2\lambda-\mu_{21}-(2\mu_{22}-5\mu_{21})n\\
    &+(-4\mu_{21}+8\mu_{22})n^2-8\mu_{22}n^3\Big)+O(\phi^3).
\end{split}
\end{equation}
As in Eq.~(\ref{eq:MSR_int}), the two fixed points at $n=0$ and $n=1$ impose $\mu_{21}=-2\lambda$. In addition, requiring the noise strength to vanish at $n=1$ fixes $\mu_{22}=\lambda$ so that $\xi[n]=\lambda(1-2n+2n^2)$. We therefore arrive at the minimal effective action, up to this order, that respects all symmetries of the path integral:
\begin{equation}\label{eq:S/N_int_PTR}
    \frac{S}{N}=\int  n\big(\partial_t+D\nabla_x^2\big)\phi+\frac{\L_{\rm int}}{N},
\end{equation}
where
\begin{equation}
    \frac{\L_{\rm int}}{N}=\lambda\phi n(1-n)(1-2n)+i\lambda \phi^2n(1-n)(1-2n+2n^2)+O(\phi^3).
\end{equation}
Remarkably, the noise strength is now directly related to the Lyapunov exponent $\lambda$, and the convergence condition~(\ref{eq:convergence}) automatically enforces $\lambda>0$! The construction of the effective action~(\ref{eq:S/N_int_PTR}) based solely on symmetry considerations constitutes the central result of this work.

We emphasize that Eq.~\eqref{eq:S/N_int_PTR} should be understood as the \textit{minimal} local action at leading derivative order and up to quadratic order in the response field. Higher-derivative terms, more general functions of $n$, and higher powers of $\phi$ are in general allowed, and the full effective action is therefore not uniquely fixed by the above argument. For example, already in the noninteracting limit~(\ref{eq:MSR_free}), there can be in principle nonlinear corrections to diffusion etc. Nevertheless, the duality condition~\eqref{eq:duality} still imposes nontrivial constraints on the structure of the effective action at higher orders.

\section{OTOC from the effective field theory}\label{sec:dynamics of OTOC}

In the previous section, we derived the general effective action $S[n,\phi]$ consistent with the symmetry conditions~\eqref{eq:normalization},~\eqref{eq:reflection},~\eqref{eq:convergence}, and the duality condition~\eqref{eq:duality}. In this section, we explain how the OTOC can be extracted from the effective action in Eq.~\eqref{eq:S/N_int_PTR}. We begin by showing that the standard OTOC, defined in Eqs.~\eqref{eq:C_1} and~\eqref{eq:C_l}, is directly represented by a two-point correlation function between $n(x,t)$ and $e^{im\phi(0,0)}$. This correlation function can then be evaluated at the saddle-point level in the path integral. Physically, the insertion of $e^{im\phi(0,0)}$ creates an initial operator of weight $m$, whose subsequent evolution is governed by the equation of motion for the density field $n(x,t)$. We show that the equation of motion following from Eq.~\eqref{eq:S/N_int_PTR} takes the form of a noisy FKPP equation. This equation supports traveling-wave solutions with exponential growth near the wavefront. The multiplicative noise has variance of order $1/N$ and gives rise to diffusive broadening after averaging over stochastic trajectories. We then support these predictions by numerical simulations of the FKPP equation and by an analytic saddle-point evaluation of the path integral.

As shown in Appendix~\ref{sec:otoc_correlator}, the OTOC between a single fermion operator at site $x$ and a time-evolved initial weight-$m$ fermionic operator at the origin, as defined in Eq.~\eqref{eq:C_l}, can be equivalently written as a two-point correlation function on the four-fold Keldysh contour:
\begin{equation}\label{eq:C_nphi}
\begin{split}
    \C_m(x,t)\equiv&\frac{1}{N^{m+1}}\sum_{j,k_1...k_m}\mathrm{Tr}\Big[\big|\{\psi_{x,j}(t),\psi_{0,k_1}...\psi_{0,k_m}\}_m\big|^2\Big]\\
    =&i\left\bra n_{s}(x,t)\left(\varphi_{s}^\hc(0,0)-\varphi_s(0,0)\right)^m\right\ket.
\end{split}
\end{equation}
Here, $n_s$ and $\varphi_s$ denote the normalized strong charge and its ladder operator, respectively, as defined in Eq.~\eqref{eq:nvarphi}. We have temporarily restored the index $s$ for clarity. The derivation proceeds by substituting the definitions of $n_s$ and $\varphi_s$ in terms of fermion bilinears, and then rewriting the resulting two-point function as a contour-ordered correlator of Majorana fields with the help of the boundary conditions~\eqref{eq:psiLpsiR|I>=0} and~\eqref{eq:psipsi|0>}. Further expressing $\varphi_s$ and $\varphi_s^\dagger$ in terms of the degrees of freedom in the EFT, we find
\begin{equation}\label{eq:C_psi^l}
\begin{split}
    \C_m(x,t)=\left \bra n(x,t)e^{im\phi(0,0)}\right\ket\left(1+O(N^{-1})\right).
\end{split}
\end{equation}
See Appendix~\ref{sec:otoc_correlator} for more details. Thus, the OTOC dynamics is fully encoded in the degrees of freedom of the effective action and can be extracted by evaluating the two-point correlation function in Eq.~\eqref{eq:C_psi^l}.

\subsection{Dynamics of the OTOC from saddle-point path integral}\label{subsec:Dynamics_OTOC}

Keeping terms up to second order in the response field $\phi$, the effective action has the general structure
\begin{equation}\label{eq:general_S}
    S[n,\phi]=N\int n\big(\partial_t+D\nabla_x^2+F[n]\big)\phi+iG[n]\phi^2+\ldots\,.
\end{equation}
The leading contribution to the two-point correlator~\eqref{eq:C_psi^l} is represented in the path-integral formalism as
\begin{equation}\label{eq:OTOC_general}
\begin{split}
    \C_{m}(x,t)=\frac{N}{Z}\int \D n_0\D\phi_0\int_{n(x,0)=n_0(x)} \D n\int_{\phi(x,0)=\phi_0(x)}\D\phi\,n(x,t)e^{im\phi_0(0)}e^{iS[n,\phi]},
\end{split}
\end{equation}
where we separate the path integral into an integral over the initial field configurations, denoted by $\phi_0(x)$ and $n_0(x)$, and an integral over field configurations subject to these initial boundary conditions. To proceed, we decouple the $\phi^2$ term by introducing a Hubbard-Stratonovich field $\xi(x,t)$, which will be interpreted as stochastic noise. After this decoupling, the action becomes linear in $\phi$, so that $\phi$ can be integrated out exactly. This imposes the saddle-point equation of motion for $n(x,t)$. Explicitly,
\begin{equation}\label{eq:OTOC_general_calculation}
\begin{split}
    \C_{m}(x,t)=&\frac{N}{Z}\int \D\xi P[\xi] \int\D n\D\phi\,\ n(x,t)e^{im\phi_0(0)}\ e^{iN\int n(\partial_t+D\nabla_x^2+F[n])\phi+\sqrt{2G[n]/N}\xi \phi}\\
    = &\frac{N}{Z}\int \D\xi P[\xi] \int \D n_0\D\phi_0\,\ n_*(n_0;x,t)e^{im\int dx\,\phi_0\delta(x)}\ e^{-iN\int dx\,n_0\phi_0}\\
    \simeq & \int \D\xi P[\xi]\,\ n_*\Big(\frac{m}{N}\delta(x);x,t\Big) 
    \equiv  \Big< n_*\Big(\frac{m}{N}\delta(x);x,t\Big)\Big>_\xi.
\end{split}
\end{equation}
In the first line, the $\phi^2$ term has been decoupled by a Gaussian field $\xi(x,t)$ with probability distribution $P[\xi] \propto {\rm exp}[-\frac{1}{2} \int dt dx \, \xi^2(x,t)]$. In the second line, we integrate over $\phi$ and $n$, which replaces the field $n(x,t)$ in the integrand by the solution $n_*(n_0;x,t)$ of the corresponding stochastic equation of motion with initial condition $n_0(x)$. The remaining integral over the initial fields imposes
\begin{equation}
n_0(x)=\frac{m}{N}\delta(x).
\label{eq:initial}
\end{equation}
Physically, this initial condition represents a weight-$m$ operator injected at the origin by the insertion $e^{im\phi(0,0)}$. We therefore arrive at the result that the OTOC is obtained by evolving the density field $n(x,t)$ according to the saddle-point equation of motion, starting from the initial condition~\eqref{eq:initial}, and then averaging over stochastic noise trajectories.

Taking the variation with respect to $\phi$, the saddle-point equation of motion takes the form
\begin{equation}\label{eq:general_FKPP}
    \partial_tn=D\nabla_x^2n+F[n]n+\sqrt{\frac{2}{N}G[n]}\xi(x,t),
\end{equation}
This is a Langevin-type stochastic reaction-diffusion equation, with deterministic growth and diffusion supplemented by multiplicative noise. The noise term is suppressed by $1/\sqrt{N}$ and therefore vanishes in the infinite-$N$ limit, where the dynamics becomes deterministic. Below, we show that for the specific effective action derived above, Eq.~\eqref{eq:general_FKPP} becomes a noisy FKPP equation.

In the following, we discuss the behavior of the OTOC predicted by the effective action~\eqref{eq:S/N_int_PTR}, using analytical and numerical approaches for the noninteracting and generic cases, respectively.

\subsection{OTOC in the noninteracting model}
\label{sec:spreading_free}

The noninteracting case is straightforward. The emergent U(1) symmetry is present, and $\lambda=0$. The equation of motion reduces to the standard diffusion equation
\begin{equation}
    \big(\partial_t-D\nabla_x^2\big)n=0\,.
\end{equation}
Since there is no noise term, the OTOC is simply the solution of this deterministic differential equation with initial condition $n(x,0)=m\delta(x)/N$, which gives
\begin{equation}\label{eq:n_*}
    \C_m(x,t)=n_*\Big(\frac{m\delta(x)}{N};x,t\Big)=\frac{m}{N\sqrt{4\pi Dt}}e^{-\frac{x^2}{4Dt}}\,.
\end{equation}
Operator spreading in noninteracting systems therefore exhibits typical diffusive behavior: a Gaussian distribution whose broadening scales as $x\sim t^{1/2}$. Numerical simulations are shown in Figs.~\ref{fig:num}(a) and~(b).

\subsection{OTOC in the interacting model}

In this case, the effective action~(\ref{eq:S/N_int_PTR}) predicts a noisy FKPP equation
\begin{equation}\label{eq:Langevin_n}
    \partial_t n-D\nabla_x^2 n=\lambda n(1-n)(1-2n)+\sqrt{\frac{2\lambda}{N}n(1-n)(1-2n+2n^2)}\xi(x,t),
\end{equation}
with Gaussian white noise satisfying $\langle \xi(x,t) \xi(x',t')\rangle =\delta(x-x') \delta(t-t')$. According to Eq.~\eqref{eq:OTOC_general_calculation}, the OTOC corresponds to the ensemble-averaged solution over noise realizations with initial condition $n(x,0)=m\delta(x)/N$. We will solve this FKPP equation numerically, discuss the deterministic and noise terms separately, and explain how the underlying discrete nature of the operator size $n$ affects the numerical simulation of the equation.

In the infinite-$N$ limit, the noise term is suppressed and the discreteness of the operator-size density $n$ can be neglected. For an initial condition localized near the origin, the early-time behavior of the OTOC is governed by the linearized equation and takes the form given in Eq.~\eqref{eq:light-cone,Lyapunov}. It exhibits exponential growth with Lyapunov exponent $\lambda$ and a ballistic light-cone structure with butterfly velocity $v_B=\sqrt{4\lambda D}$. The nonlinear terms in $n$ then stop the indefinite exponential growth and determine the saturation regime. In particular, inside the ballistic light cone, the local growth of the OTOC slows down and the operator-size density approaches the stable fixed point $n=1/2$.

At large but finite $N$, two additional effects become important. First, the multiplicative noise term in the FKPP equation is restored and leads to visible front fluctuations in numerical simulations. Second, when $Nn\sim O(1)$, the discreteness of the underlying operator-size degrees of freedom can no longer be ignored. In the continuum FKPP description, this discreteness is commonly modeled by introducing an effective cutoff.

The physical origin of the cutoff is as follows. In a spatially local model, neither deterministic growth nor stochastic noise should nucleate operator weight in a region where the operator is completely absent. This property is already reflected in Eq.~\eqref{eq:Langevin_n}: both the deterministic growth term and the noise amplitude vanish at $n=0$. At finite $N$, however, the continuum description is meaningful only once the local operator weight corresponds to at least one microscopic degree of freedom, namely when $Nn\gtrsim 1$. We therefore implement discreteness effects by multiplying both the deterministic and noise terms by an effective hard cutoff $\theta(n-1/N)$. This prescription prevents the continuum FKPP equation from artificially amplifying densities below the microscopic scale set by $1/N$. 

\begin{figure}[t]
    \centering
    \includegraphics[width=1\linewidth]{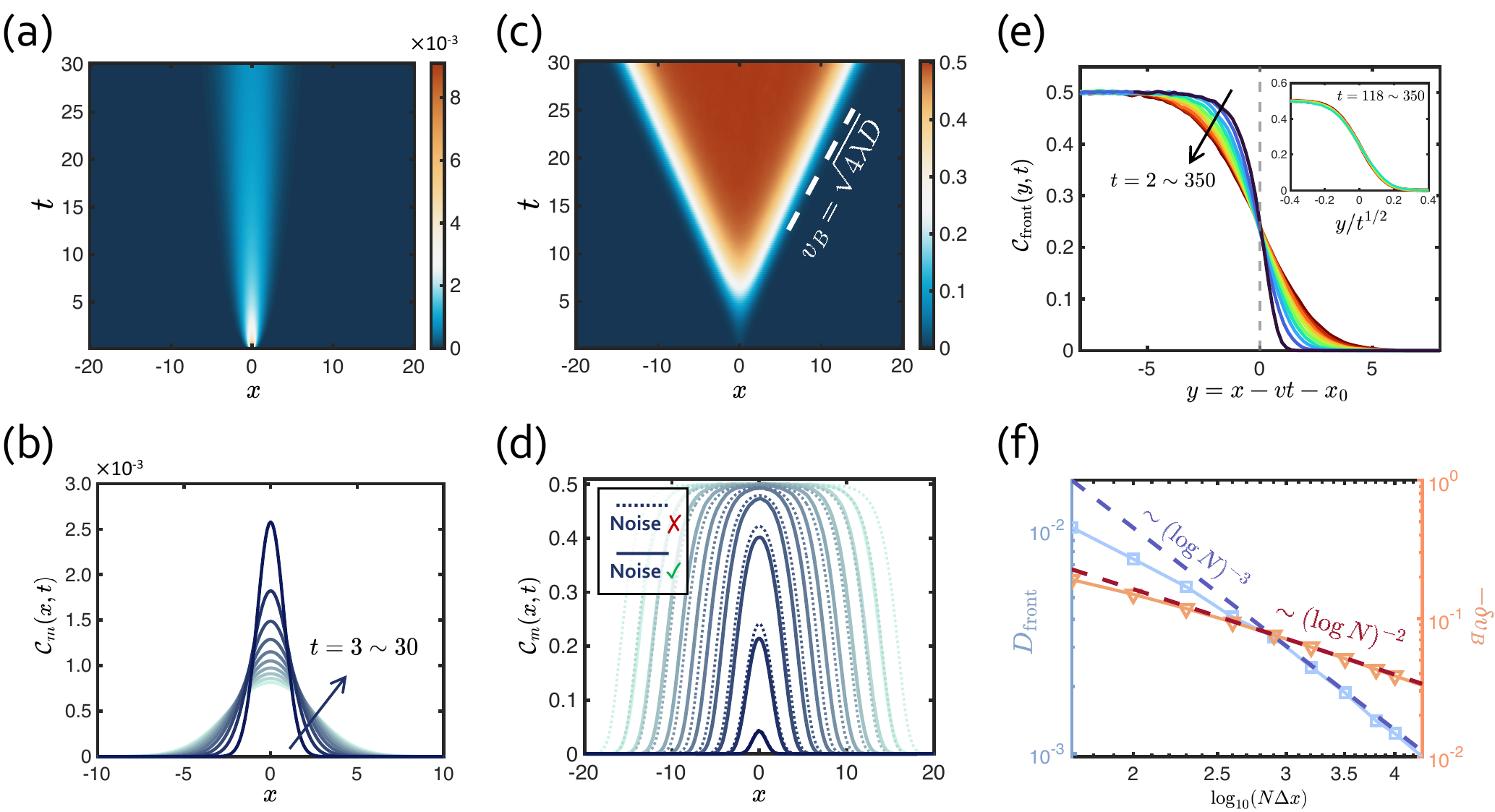}
    \caption{Numerical solutions of the FKPP equation \eqref{eq:Langevin_n}, starting from an initial operator at the origin with $m=5$, for (a)\&(b) the noninteracting case with $\lambda=0$, and (c)\&(d) the interacting case with $\lambda=1$, averaged over $10^3$ samples. Time slices for $t=3\sim 30$ are shown in (b) and (d), respectively. In panel (d), the solid curves denote the noisy solutions, while the dashed curves denote the solutions of the deterministic equation without noise. The numerical parameters are chosen as $D=0.1$ and $N\Delta x=10^3$. Noise and the $1/(N\Delta x)$ cutoff induce (c) a reduction of the butterfly velocity and (e) wavefront broadening $\propto t^{1/2}$. In panel (e), the FKPP equation is solved numerically using a fully developed wavefront as the initial condition, yielding a ballistic front with diffusive broadening. (f) shows the scaling of wavefront diffusion constant $D_{\rm front}$ and the reduction of the butterfly velocity, $-\delta v_B$ with $\log N$ (dashed line). For computational efficiency, the parameters in (e) and (f) are chosen as $D=0.04$.}
    \label{fig:num}
\end{figure}

With this cutoff prescription, the stochastic FKPP equation~\eqref{eq:Langevin_n} can be solved numerically. The result is shown in Fig.~\ref{fig:num}(c)\&(d). Comparing the solid curve, obtained with noise, to the dashed curve, obtained with the noise term turned off, reveals one important effect of finite-$N$ fluctuations: noise slows down the averaged growth of the OTOC. This effect can be understood directly from Eq.~\eqref{eq:Langevin_n}. To compare with deterministic evolution, we average the stochastic equation over noise realizations. The explicit noise term drops out, and the left-hand side is unchanged because it is linear in $n$. The difference comes from the nonlinear deterministic growth term. Writing
\begin{equation}
n=\langle n\rangle_{\xi}+\delta n,
\qquad
\langle \delta n\rangle_{\xi}=0,
\end{equation}
and expanding the reaction term $\lambda n(1-n)(1-2n)$ around $\langle n\rangle_{\xi}$, the leading correction due to fluctuations is
\begin{equation}
-3\lambda\bigl(1-2\langle n\rangle_{\xi}\bigr)
\langle(\delta n)^2\rangle_{\xi}.
\end{equation}
In the growing regime, where $0<\langle n\rangle_{\xi}<1/2$, this contribution is negative. Thus fluctuations reduce the effective growth rate of the noise-averaged OTOC. This explains why the stochastic evolution grows more slowly than the deterministic evolution, especially at intermediate times when the front has developed appreciable fluctuations but has not yet fully saturated.

In addition to slowing down the averaged growth, finite-$N$ noise and the cutoff lead to two further effects: broadening of the operator front and a reduction of the butterfly velocity. The front may be identified with the leading edge of the ballistic light cone. For instance, the right-moving edge of the linearized solution in Eq.~\eqref{eq:S/N=lambda} has the form $\sim e^{-2\lambda z}$, with $z=x/v_B-t$. More generally, for sufficiently localized initial conditions, the deterministic large-$N$ FKPP equation selects a sharp pulled front whose leading edge decays as $e^{-\gamma z}$. This picture is modified once finite-$N$ discreteness and noise are included. In the theory of noisy FKPP fronts, Brunet {\it et al.}~\cite{brunet1997shift,brunet2006phenomenological} showed that an effective cutoff at a density of order $1/N$ reduces the asymptotic front velocity relative to its deterministic value. Moreover, the front position itself also becomes a fluctuating quantity and undergoes diffusion due to the presence of noise. At large $N$, the corresponding velocity shift and front-diffusion constant scale as
\begin{equation}
\delta v_B\sim -(\log N)^{-2},
\qquad
D_{\rm front}\sim (\log N)^{-3}.
\end{equation}
Our numerical simulations of the cutoff noisy FKPP equation, initialized with a fully developed front, reproduce these effects. The broadening of the front and the reduction of the butterfly velocity are shown in Figs.~\ref{fig:num}(e) and (c), respectively, while the scaling with $N$ is shown in Fig.~\ref{fig:num}(f). The resulting noise-averaged front is no longer characterized by a purely exponential leading edge.

\section{Example: the Brownian SYK model}\label{sec:SYK}

\begin{figure}[t]
    \centering
    \includegraphics[width=0.8\linewidth]{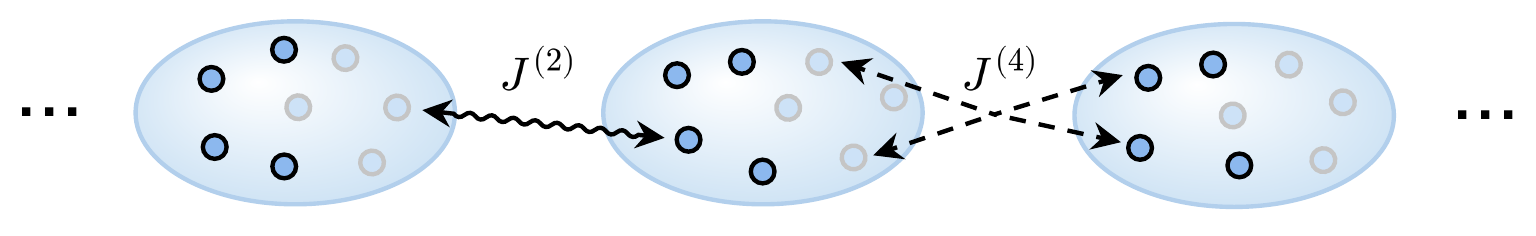}
    \caption{Schematic illustration of the Brownian SYK model. The Hamiltonian, defined in Eq.~\eqref{eq:sec6:Ham}, contains both nearest-neighbor two-body and four-body interactions.}
    \label{fig:SYK_cartoon}
\end{figure}

We now consider a concrete example, the Brownian SYK chain, for which the effective action governing the OTOC can be derived explicitly. The model consists of a one-dimensional chain with $N$ Majorana fermions on each site. We include both nearest-neighbor quadratic hopping and nearest-neighbor four-body interactions:
\begin{equation}\label{eq:sec6:Ham}
\begin{split}
    H(t)=\frac{i}{2}\sum_{x,jk}J_{x,jk}^{(2)}(t)\psi_{x,j}\psi_{x+1,k}-\frac{1}{4!}\sum_{x,jklm}J_{x,jklm}^{(4)}(t) \psi_{x,j}\psi_{x,k}\psi_{x+1,l}\psi_{x+1,m}.
\end{split}
\end{equation}
The first term describes quadratic hopping between adjacent sites, while the second term describes quartic interactions involving Majorana fermions on adjacent sites. A schematic illustration of the model is shown in Fig.~\ref{fig:SYK_cartoon}. The time-dependent couplings are Gaussian variables with mean values $\langle J^{(2)}\rangle=\V$, $\langle J^{(4)}\rangle=0$ and variances
\begin{equation}
    \left\bra J_{x,jk...}^{(q)}(t)J_{x',j'k'...}^{(q')}(t')\right\ket=\frac{q!\J_q}{N^{q-1}}\delta(t-t')\delta^{qq'}\delta_{xx'}\delta_{jj'}\delta_{kk'},
\end{equation}
for $q=2,4$. Here $\mathcal J_q$ controls the strength of the Brownian $q$-body coupling. We have allowed a nonzero mean $\V$ in the quadratic coupling in order to include a uniform hopping component, whose effects will be discussed separately in Sec.~\ref{subsec:SYK_clean}.

In the remainder of this section, we first derive the quadratic effective action using the saddle-point expansion in Sec.~\ref{subsec:SYK_action_2}. We then extend the derivation and obtain the effective action to all orders in Sec.~\ref{subsec:SYK_action_4}. Finally, in Sec.~\ref{subsec:SYK_clean}, we discuss the special case of a translationally invariant free-fermion system. In the main text, we only outline the derivation and summarize the key results; the detailed calculations are presented in Appendix~\ref{appsec:SYK}.

\subsection{Effective action at quadratic order}\label{subsec:SYK_action_2}

We first rewrite Hamiltonian~\eqref{eq:sec6:Ham} in terms of complex fermionic operators in the entangled $L/R$ basis~\eqref{eq:entangle} on the $+$ and $-$ contours. Using the notation $(c^{1},c^2,c^3,c^4)\equiv(c^{+\dagger},c^+,c^{-\dagger},c^-)$, the partition function on the Keldysh contour is
\begin{equation}
    Z=\int \prod_{\a=1}^4\D c^\a\, \exp\left\{i\int_t\sum_{\b=1}^4\sum_{x,j} \frac{1}{2}c_{x,j}^\b\partial_tc_{x,j}^{\bar{\b}}-H_Q[\{c^\a\};t]\right\},
\end{equation}
where $c^{\bar{\b}}\equiv(c^\b)^\dagger$ such that $\bar 1=2$, $\bar 3=4$ and vice versa, and $\int_t$ is shorthand for integration over time. To integrate out the fermionic fields, we proceed as in standard SYK models by introducing the fermion bilinears
\begin{equation}\label{eq:G_def}
    G_x^{\a\b}(t_1,t_2)=\frac{1}{N}\sum_{j=1}^{N} c_{x,j}^\a(t_1)c_{x,j}^{\b}(t_2)\,,
\end{equation}
and the corresponding self-energy $\Sigma^{\a\b}_x(t_1,t_2)$ enforcing the constraint
\begin{equation}
    \delta\Big(G_x^{\a\b}- \frac{1}{N}\sum_jc^{\a}_{x,j}c^{\b}_{x,j}\Big)=\int\D\Sigma\, e^{iN\int_{t_1t_2} \Sigma^{\a\b}_x\big(G_x^{\a\b}-\frac{1}{N}\sum_j c^{\a}_{x,j}c^{\b}_{x,j}\big)}.
\end{equation}
Here, we suppress the temporal dependence for notational simplicity. By definition, the self-energy is antisymmetric, $\Sigma^{\a\b}=-\Sigma^{\b\a}$. Integrating out the Gaussian variables $J^{(2)}$ and $J^{(4)}$, as well as the complex fermionic fields $\{c^{\a}\}$, we obtain the action for the bilocal fields
\begin{equation}\label{eq:sec6_effaction}
    \begin{split}
        \frac{iS[\Sigma,G]}{N}=&\log \Pf\Big[\frac{1}{2}\sigma^0\otimes \sigma^x(\partial_t-2i\V\sin{k})+\Sigma\Big]\\
        &+\sum_{\a,\b\neq\a}\sum_x\int_{t_1t_2} G^{\a\b}_x\Sigma^{\a\b}_x-\Big(\frac{\J_2}{4} G^{\a\b}_x G^{\bar{\a}\bar{\b}}_{x+1}
        +\frac{\J_4}{8} F^{\a\b}_x(G)\Big)\delta(t_{12}),
    \end{split}
\end{equation}
where $t_{12}=t_1-t_2$, and $F^{\a\b}_x(G)$ is a quartic polynomial of $G^{\a\b}_x$ whose explicit form is given in Appendix~\ref{appsec:SYK}.

Since the action $S[\Sigma,G]$ is proportional to $N$, we perform a saddle-point expansion in the large-$N$ limit. The saddle point is given by solutions of the Schwinger-Dyson equations for the $G$ and $\Sigma$ fields subject to the appropriate boundary conditions. In our case, the saddle-point solution reads
\begin{equation}
    \overline{G}(t)=e^{-\Gamma|t|}J_0(2\V|t|)\begin{pmatrix}
        0 & -\theta(-t)& -i & 0\\\theta(t) & 0&0&0\\
        i&0&0&-\theta(-t)\\0&0&\theta(t)&0
    \end{pmatrix},
\end{equation}
where $\Gamma=\J_2/2+\J_4/4$, and $J_0(x)$ is the Bessel function. Details of solving the saddle point equation are presented in Appendix~\ref{appsec:SYK}.

Next we consider fluctuations around this saddle point:
\begin{eqnarray}
    G^{\a\b}_x(t_1,t_2)&=&\overline{G}^{\a\b}(t_{12})+\delta G_x^{\a\b}(t_1,t_2), \\
    \Sigma^{\a\b}_x(t_1,t_2)&=&\overline{\Sigma}^{\a\b}(t_{12})+\delta \Sigma_x^{\a\b}(t_1)\delta(t_{12}). 
\end{eqnarray}
In principle, the action~\eqref{eq:sec6_effaction} can be expanded to arbitrary order in these fluctuations. The zeroth-order term is simply the saddle-point value, while the first-order term vanishes by definition. We therefore focus on the quadratic and higher-order terms. At quadratic order in $\delta \Sigma$ and $\delta G$, we can integrate out the self-energy fluctuations, which imposes the constraint $\delta G^{12}=\delta G^{34}=i\delta G^{24}$. The quadratic effective action for the remaining degrees of freedom is
\begin{equation}\label{eq:sec6_eff_deltaG}
        \frac{S_{\rm eff}}{N}=\int_{k,\omega} \delta G^{12}_{-k}(-\omega ) \big[-i\Omega_{k,\omega}+(2\Gamma+\J_4)\cos{k}\big]\big(\delta G^{13}_k(\omega)-i\delta G_k^{12}(\omega)\big),
\end{equation}
where
\begin{equation}\label{eq:f,V}
    \Omega_{k,\omega}=\left((\omega-2i\Gamma)^2-16\V^2\sin^2(k/2)\right)^{1/2}.
\end{equation}
Here and below, we denote $\int_{k,\omega}=\int dk\,d\omega/(2\pi)^2$, and $\O_k(\omega)=\int_{k,\omega}e^{i\omega t-ikx}\O_x(t)$.

To make contact with the general EFT constructed in terms of the $(n,\phi)$ degrees of freedom, we express the bilocal fields in terms of the operator-size field introduced earlier and its corresponding ladder operators. The connection is straightforward:
\begin{equation}\label{eq:G_nphi}
    \frac{1}{2}\big(G^{12}+G^{34}\big)=n_s,\quad G^{24}=-\vphi_s,\quad G^{13}=\vphi_s^\hc\,.
\end{equation}
Therefore, substituting Eq.~\eqref{eq:delta_n_delta_phi} and using $\delta\O_x=\O_x-\bra \O\ket$, the linear fluctuations of these bilocal fields are $\delta G^{12}=\delta G^{34}=n$, $\delta G^{13}=in+\phi$, and $\delta G^{24}=-in$, which satisfy the constraint obtained by integrating out $\delta\Sigma$. In terms of the new degrees of freedom, the effective action becomes
\begin{equation}\label{eq:sec6_eff_nphi}
    \begin{split}
        \frac{S_{\rm eff}}{N}=&\int_{k,\omega}n_{-k}(-\omega)\big[-i\Omega_{k,\omega}+(2\Gamma+\J_4)\cos{k}\big]\phi_k(\omega)\\
        \approx &\int_{x,t}n\big(\partial_t+D\partial_x^2+\J_4\big)\phi\,,
    \end{split}
\end{equation}
where $D=\Gamma(1+\V^2/\Gamma^2)+\J_4/2$. In the second line, we take the hydrodynamic limit $k,\omega\to 0$ and retain only the leading terms. Thus, at quadratic order, we reproduce the effective action~\eqref{eq:S/N=lambda} without the noise term, which captures the early-time exponential growth of the OTOC. In particular, the Lyapunov exponent governing this early-time growth depends only on $\mathcal J_4$. If the interaction term is turned off, namely $\mathcal J_4=0$, the OTOC instead exhibits diffusive behavior. This confirms that the interaction term breaks the emergent U(1) symmetry and thereby generates a mass term for the diffusive Goldstone mode at lowest order.

\subsection{Effective action beyond the quadratic order}\label{subsec:SYK_action_4}

To obtain the higher-order effective action, two ingredients must be taken into account. First, higher-order terms in $\delta\Sigma$ may violate the quadratic-order constraint $\delta G^{12}=i\delta G^{24}$, and therefore require a modification of the kernel term in the effective action. Second, the interaction term can contain arbitrary powers of $\delta G$, among which the relevant higher-order contributions are the on-site terms. At the highest order, the effective action is derived in Appendix~\ref{appsec:SYK} as
\begin{equation}\label{eq:Seff_higher_G}
    \begin{split}
        \frac{S_{\rm eff}}{N}=&\int_{k,\omega}\big[\Omega_{k,\omega}+(2i\Gamma+i\J_4)\cos{k}\big]\,\big[\delta G^{24}_{-k}(-\omega)\delta G^{13}_k(\omega)-\big|\delta G^{12}_k(\omega)\big|^2\big]\\
        &+\frac{i\J_4}{2}\int_{x,t}4(\delta G^{12}_x)^3+3i(\delta G^{13}_x)^2\delta G^{24}_x+i(\delta G^{24}_x)^3\\
        &-\frac{i\J_4}{2}\int_{x,t}2(\delta G^{12}_x)^4+(\delta G^{13}_x)^3\delta G^{24}_x+\delta G^{13}_x(\delta G^{24}_x)^3.
    \end{split}
\end{equation}
Here the kernel term is not derived from integrating out $\delta\Sigma$ exactly to all orders, which is technically impossible. Instead, it is obtained by a reasonable modification of the quadratic kernel such that it reduces to Eq.~\eqref{eq:sec6_eff_deltaG} at quadratic order and ensures that the symmetry breaking terms at arbitrary order receive no contribution from $\J_2$. Using the identification in Eq.~\eqref{eq:G_nphi}, the above fluctuations can be expressed in terms of the $(n,\phi)$ degrees of freedom as
\begin{equation}
    \delta G^{12}=\delta G^{34}=n,\quad \delta G^{13}=-i(1-n)e^{i\phi}+i,\quad \delta G^{24}=-ine^{-i\phi},
\end{equation}
where we substitute the full form of Eq.~\eqref{eq:delta_n_delta_phi}, rather than only its linear order. Substituting this into the effective action~\eqref{eq:Seff_higher_G}, then retaining terms up to order $\phi^2$ and taking the hydrodynamic limit $k,\omega\to 0$ while keeping the relevant derivative terms, we finally obtain
\begin{equation}
    \frac{S_{\rm eff}}{N}=\int_{x,t} n\big(\partial_t+D\partial^2_x\big)\phi+\J_4 n(1-n)(1-2n)\phi+i\J_4 n(1-n)(1-2n+2n^2)\phi^2+...,
    \label{eq:S_SYK}
\end{equation}
which is precisely the effective action~\eqref{eq:S/N_int_PTR} obtained from general symmetry constraints, with Lyapunov exponent $\lambda=\J_4$.

\begin{figure}[t]
    \centering
    \includegraphics[width=0.95\linewidth]{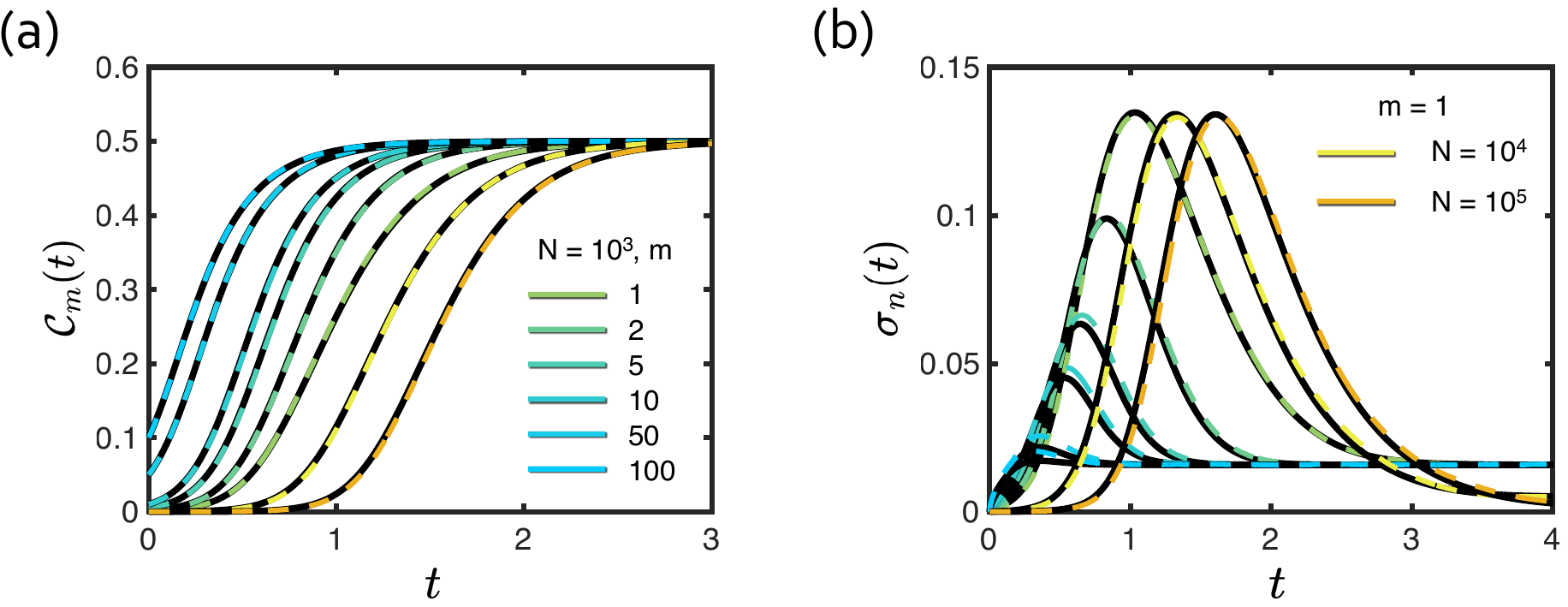}
    \caption{Numerical simulations of the $(0+1)d$ FKPP equation predicted by~(\ref{eq:S_SYK}) (dashed color lines) and comparison with the operator size computed from the master-equation approach in the Brownian SYK$_4$ model~\cite{agarwal2022emergent} (black solid lines). 
    The first moment of $n(t)$ corresponds to the OTOC, and is shown in (a). The standard deviation $\sigma_n=(\langle n^2\rangle-\langle n\rangle^2)^{1/2}$, related to the second moment of $n$, is shown in (b).
    The parameters are chosen as $\lambda=\J_4=8$, the number of simulation samples is $5\times 10^4$, and the finite-$N$ cutoff is taken to be $\theta(n-m/N)$.}
    \label{fig:(0+1)dFKPP}
\end{figure}

The probability distribution of the OTOC in the Brownian SYK model can also be computed numerically using the master equation derived in Refs.~\cite{agarwal2022emergent,yao2024notes,xu2025dynamics}. The main idea is to integrate out the random couplings to obtain an averaged Hamiltonian whose individual terms can all be written as combinations of generators of an $su(2)\oplus su(2)$ algebra. This structure, together with other symmetries, reduces the Hilbert space to an $N^{M^d}$-dimensional subspace spanned by irreducible representations of the $su(2)\oplus su(2)$ algebra, and leads to the master equation governing the dynamics of the probability distribution.

Owing to the large-$N$ nature of the SYK model, the $N^{M^d}$-dimensional Hilbert space remains very difficult to simulate. However, for the single-site model in $d=0$, the master equation can be simulated efficiently. In Fig.~\ref{fig:(0+1)dFKPP}, we compare the (0+1)d FKPP equation derived from our effective action~(\ref{eq:S_SYK}) and the master equation derived in~\cite{agarwal2022emergent}. The numerical results for the first two moments of the operator size $n(t)$ show excellent agreement with the master equation approach.
Notice that in this case the diffusion term in the FKPP equation is absent, and we adopt the noise cutoff $\theta(n-m/N)$ for initial operator of weight $m$. 
A more direct verification of our general theory is that, after taking the continuum limit of the master equation, the differential equation for the corresponding probability distribution can be mapped to an Ito process~\cite{yao2024notes}, whose associated FKPP equation is exactly identical to the $(0+1)$d FKPP equation predicted by~(\ref{eq:S_SYK}).

\begin{figure}[t]
    \centering
    \includegraphics[width=1\linewidth]{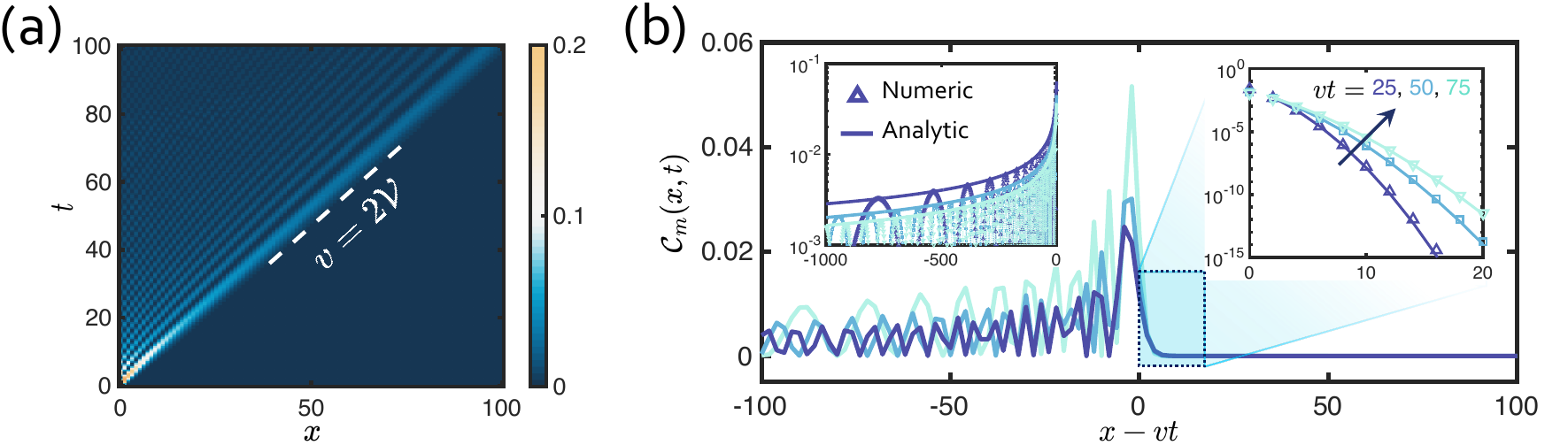}
    \caption{(a) OTOC for the translationally invariant SYK$_2$ chain Eq.~(\ref{eq:sec6_int_C}) with $\V=0.5$. A light-cone emerges with velocity $v=2\V$ and the OTOC does not saturate inside the light-cone. (b) Horizontal cuts of (a) at $vt=25,50,75$. The insets show zoom-in views of the behaviors deep inside the light-cone and near the front. Left: $\C_m(x,t)$ fluctuates deep inside the light-cone, with an envelope decaying according to Eq.~(\ref{eq:sec6_tail}). Right: near the front, $\C_m(x,t)$ follows Eq.~(\ref{eq:sec6_wf_exp}), with a broadening $\propto t^{1/3}$.}
    \label{fig:clean}
\end{figure}

\subsection{SYK$_2$ chain with translation invariance}\label{subsec:SYK_clean}

Interestingly, we can turn off the temporal fluctuations $\mathcal J_2$ and $\mathcal J_4$ in Eq.~\eqref{eq:sec6_eff_nphi}. Since the hopping term has a nonzero mean, it survives in this limit, which corresponds to a static, translationally invariant SYK$_2$ chain. In this case, the effective action is no longer local because of the square-root factor, and therefore lies beyond the form predicted by our general EFT. This nonlocality results from integrating out the slow modes associated with additional conservation laws, namely the infinitely many conserved densities in the clean model. Nevertheless, the OTOC can still be computed as
\begin{equation}\label{eq:sec6_int_C}
\begin{split}
    \C_m(x,t)&=\frac{N}{Z}\int \D n\D\phi \,n (x,t)e^{iN\int_{k,\omega}n_{-k}(-\omega)(-i\Omega_{k,\omega})\phi_k(\omega)+\frac{m}{N}\phi_k(\omega)}\\
    &\approx  im\int_{k,\omega}\frac{e^{i(kx-\omega t)}}{\sqrt{\omega^2-16\V^2\sin^2(k/2)}}
    =\frac{m}{2}\int _kJ_0\big(4\V t\left|\sin(k/2)\right|\big)e^{ikx}.
\end{split}
\end{equation}
The numerical result of the final Fourier transform is shown in Fig.~\ref{fig:clean}(a), exhibiting a linear light-cone with velocity $v=2\V$. However, because of the integrability of the clean model, there is neither exponential growth at the front nor saturation of the OTOC inside the light-cone. Figure~\ref{fig:clean}(b) shows slices along the edge of the light-cone, displaying oscillations in the bulk of the light-cone as well as a distinct scaling behavior of the front.

In the late-time limit, the scaling behavior of the wavefront can be computed analytically. By expanding the $k$ dependence in Eq.~\eqref{eq:sec6_int_C} up to the subleading $k^3$ term, the asymptotic behavior of the OTOC at the wavefront on the $x>0$ branch is found in Appendix~\ref{appsec:clean} to be
\begin{equation}\label{eq:sec6_wf_exp}
    \C_m(x,t)\sim\frac{m}{\sqrt{2vt(x-vt)}}e^{-\frac{4\sqrt{2}}{3}\frac{(x-vt)^{3/2}}{(vt)^{1/2}}},
\end{equation}
where the light-cone velocity is $v=2\mathcal V$, as predicted numerically. This scaling behavior with $(x-vt)\sim t^{1/3}$ is compared with the exact numerical integral in the inset of Fig.~\ref{fig:clean}(b), which shows excellent agreement. The same scaling behavior was also noted in Refs.~\cite{khemani2018velocity,xu2020accessing,xu2024scrambling}. It belongs to a universality class distinct from those of genuinely interacting large-$N$ models. In addition, the asymptotic behavior of the oscillatory tail deep inside the light-cone is
\begin{equation}\label{eq:sec6_tail}
    \C_m(x,t)\sim \frac{m}{2}\int_k J_0\big(2\V t|k|\big)e^{ikx}= \frac{m}{\pi\sqrt{v^2t^2-x^2}},
\end{equation}
which also shows excellent agreement with the numerical integral [see inset of Fig.~\ref{fig:clean}(b)].

\section{Summary and discussion}

We construct an effective field theory for fermionic operator scrambling from the symmetry structure of the OTOC contour. 
In the doubled-Hilbert-space representation of the four-fold Keldysh contour, quadratic Majorana dynamics exhibits an emergent strong U(1) symmetry. The corresponding charge counts operator size, while its conjugate phase plays the role of the response field in the Schwinger-Keldysh effective action. In the absence of interactions, the resulting theory describes diffusive spreading of operator size, consistent with the behavior of random free fermions.

Interactions with $q\geq 4$ explicitly break the emergent U(1) symmetry and allow terms that are forbidden in the quadratic limit. In the effective theory, this explicit breaking appears as a mass term for the would-be Goldstone mode. Its linear coefficient gives the Lyapunov exponent governing the early growth of the OTOC. Together with the diffusion constant of the operator-size density, this growth rate determines the butterfly velocity. Going beyond the linearized theory, we show that the nonlinear terms are strongly constrained by a microscopic symmetry that combines time reversal with a permutation of the Keldysh contours, which fixes the minimal action up to order $\phi^2$. In particular, it relates the multiplicative noise strength to the same parameter that controls the deterministic Lyapunov growth, which naturally enforces the positivity of the Lyapunov exponent.

The OTOC is then obtained by inserting an initial operator-size source and evolving the operator-size density with the stochastic dynamics generated by the effective action. This leads to a noisy FKPP-type equation. The equation contains the expected leading-edge exponential growth and ballistic propagation, but also encodes the late-time saturation behavior and finite-$N$ front fluctuations. We further check our construction in a Brownian SYK chain, where a direct saddle-point expansion reproduces the form of the effective action derived from the symmetry constraints.

The main lesson is that the reaction-diffusion description of scrambling need not be introduced as a phenomenological master equation. In the class of large-$N$ fermionic systems considered here, the deterministic growth term, the saturation structure, and the leading Gaussian noise follow from Schwinger-Keldysh consistency, the operator-space interpretation of the doubled contour, and the time reversal/contour-permutation duality together provide a symmetry-based route to operator size hydrodynamics, in which the relevant slow variable is not an ordinary conserved density but an emergent degree of freedom associated with operator evolution.

Several questions remain open. One natural extension is to include ordinary conserved quantities, such as energy or charge. In that case, the operator-size field should couple to conventional hydrodynamic modes, and the resulting theory should describe the interplay between scrambling, transport, and long-time tails in OTOCs. Another direction is to incorporate finite-temperature effects. This would require imposing the appropriate KMS constraints and may modify both the effective action and the relation between operator size and OTOCs.

It would also be useful to understand how far the present construction extends beyond the large-$N$ fermionic setting. Spin and bosonic systems have different operator algebras and different saturation structures, so their nonlinear effective theories need not take the same form as the fermionic FKPP equation derived here. Similarly, higher-order terms in the response field, which encode non-Gaussian noise and rare front fluctuations, remain to be systematically constrained. Clarifying these issues may lead to a broader EFT framework for quantum scrambling in deterministic Hamiltonian systems, random circuits, long-range interacting models, and holographic theories.

\section*{Acknowledgments}

B.-L.C. and Z.-C.Y. are supported by Grant No. 12375027 from the National Natural Science Foundation of China.
The work of S.-K. J. is supported by a start-up fund at Tulane University.

\newpage
\appendix

\section{Expressing SU(2) generators via $n,\phi$}\label{app:nphi_parameterization}

From first principles, in a coherent-state or collective-field path integral, operators are represented by their matrix elements in a complete set of intermediate states. These matrix elements define complex functions of the path-integral variables, which we refer to as classical symbols.

In this appendix, we explain the parametrization of the classical symbols associated with the operators $\mathcal N_\sigma/N$, $\Phi_\sigma/N$, and $\Phi^\dagger_\sigma/N$, with $\sigma=s,w$, which generate an emergent $su(2)\oplus su(2)$ algebra. In terms of $n=\N/N$, $\vphi=\Phi/N$ and $\vphi^\hc=\Phi^\hc/N$, the algebra is
\begin{equation}
    [n_{s,x},\vphi_{s,y}]=-\frac{1}{N}\vphi_{s,x}\delta_{xy},\quad \mathrm{h.c.},\quad [\vphi_{s,x}^\hc,\vphi_{s,y}]=\frac{1}{N}(2n_{s,x}-1);
\end{equation}
\begin{equation}
    [n_{w,x},\vphi_{w,y}]=-\frac{2}{N}\vphi_{w,x}\delta_{xy},\quad \mathrm{h.c.},\quad [\vphi_{w,x}^\hc,\vphi_{w,y}]=\frac{1}{N}n_{w,x};
\end{equation}
with $[\O_{s,x},\O'_{w,y}]=0$. The expectation values relevant for the OTOC generating functional are transition matrix elements, $\bra...\ket=\bra I|...|O\ket$, and at the saddle they obey
\begin{equation}\label{appeq:<>}
    \big< \vphi_s^\hc\big>=-i,\quad \bra n_s\ket=\bra \vphi_s\ket=0;\qquad \bra n_w\ket=\bra \vphi_w\ket=\big< \vphi_w^\hc\big>=0\,.
\end{equation}
Moreover, the initial state is an eigenstate of the corresponding Casimir operators:
\begin{equation}
    \frac{ J^2_\sigma|O\ket}{N^2}=\left(\frac{1}{4}+\frac{1}{2N}\right)\delta_{\sigma s}|O\ket\,.
\end{equation}

For the strong sector, the leading large-$N$ saddle is therefore associated with a nonzero fixed-Casimir orbit. After projecting to this saddle orbit, the path integral may be described by spin coherent-state variables in the $j=N/2$ sector, and the normalized generators are replaced by their coherent-state classical symbols. This statement should be understood as a leading low-energy description around the saddle, rather than as an exact restriction of the full microscopic dynamics to one Casimir sector.

The weak sector, however, requires a different interpretation. Since the corresponding saddle has zero Casimir, it should not be identified with an ordinary spin coherent-state orbit at fixed $j=0$, because that orbit is trivial and would give vanishing symbols for all $su(2)$ generators. Instead, the weak variables describe fluctuations around a zero-Casimir saddle in the full collective Hilbert space. The associated collective symbols are constrained by the algebra and by the transition matrix elements, and they can be parametrized by a complexified zero-Casimir (or nilpotent) orbit. In this parametrization, $n_w$ itself is the amplitude variable emerging from the zero saddle.

Thus, after imposing the relevant saddle constraints, two canonical coordinates are sufficient in each sector. We parametrize the admissible classical symbols as
\begin{equation}\label{appeq:nphi_expansion_1}
    \begin{cases}
        n_s=n_s,\\
        \vphi_s=if_s[n_s]e^{-i\phi_s},\\
        \vphi_s^\hc=-ig_s[n_s]e^{i\phi_s},
    \end{cases}\quad \begin{cases}
        n_w=n_w,\\
        \vphi_w=f_w[n_w]e^{-2i\phi_w},\\
        \vphi_w^\hc=g_w[n_w]e^{2i\phi_w},
    \end{cases}
\end{equation}
where the coordinates are chosen as two canonical pairs,
\begin{equation}
    [\phi_{\sigma,x},n_{\rho,y}]=\frac{i}{N}\delta_{\sigma\rho}\delta_{xy},\quad \sigma,\rho\in\{s,w\}.
\end{equation}
With this choice, the ansatz automatically reproduces the $[n,\vphi]$-type commutators. The saddle expectation values \eqref{appeq:<>} require $f_s[0]=f_w[0]=g_w[0]=0$ and $g_s[0]=1$. To further determine the form of these functions, we impose the conditions from the $[\varphi^\dagger,\varphi]$-type commutators, which give
\begin{equation}
    f_s\Big[n-\frac{1}{N}\Big]\cdot g_s[n]-f_s[n]\cdot g_s\Big[n+\frac{1}{N}\Big]=\frac{1}{N}(2n-1),
\end{equation}
\begin{equation}
    f_w\Big[n-\frac{2}{N}\Big]\cdot g_w[n]-f_w[n]\cdot g_w\Big[n+\frac{2}{N}\Big]=\frac{1}{N}n.
\end{equation}

For the strong sector, expanding in powers of $1/N$ gives
\begin{equation}
    f_s'\cdot g_s+f_s\cdot g_s'=1-2n;\qquad f^{(k)}_s\cdot g_s-(-1)^kf_s\cdot g_s^{(k)}=0,\qquad \forall k>1\,,
\end{equation}
which together with $f_s[0]=0$ and $g_s[0]=1$ restrict that $f_s[n]=n$ and $g_s[n]=1-n$. Similarly, the solutions of the weak sector are $f_w[n]=ic^{-1}n/2$ and $g_w[n]=icn/2$. The phase of $c$ can be absorbed into $\phi_w$, while its modulus is a normalization convention for the complexified nilpotent coordinates. We choose the symmetric normalization $c=1$. The final parametrization is therefore
\begin{equation}
    \begin{cases}
        n_s=n_s,\\
        \vphi_s=in_se^{-i\phi_s},\\
        \vphi_s^\hc=-i(1-n_s)e^{i\phi_s},
    \end{cases}\quad \begin{cases}
        n_w=n_w,\\
        \vphi_w=\frac{i}{2}n_we^{-2i\phi_w},\\
        \vphi_w^\hc=\frac{i}{2}n_we^{2i\phi_w}.
    \end{cases}
\end{equation}
One can verify that the classical Casimir obtained from this parametrization satisfies $J_{\sigma,\rm cl}^2/N^2=\delta_{\sigma s}/4$.

\section{Constraints from Schwinger-Keldysh formalism}\label{app:SK_constraints}

In this section, we present proofs of several constraints in the main text, which mainly follows Ref.~\cite{glorioso2016second}.

In the Schwinger-Keldysh path integral formalism, the partition function is given by integrating out fast modes $\psi_\a\in\{\psi_+,\psi_-\}$ and slow modes $\phi_\a\in\{\phi_+,\phi_-\}$, which reads
\begin{equation}
    Z=\int_{\rho_0}\D\phi_\pm\D\psi_\pm e^{iS_0[\psi_+,\phi_+]-iS_0[\psi_-,\phi_-]}.
\end{equation}
Here, we simply denote $\D\chi_\pm=\D\chi_+\D\chi_-$, $\rho_0$ refers to the initial density matrix which depends on the initial configurations of fields, which are denoted as $\phi_\a^0$ and $\psi_\a^0$ with $\a=\pm$. We then extract the integrand functional for the integral over $\phi_\a$, which is
\begin{equation}\label{eq:app_Z=intDA}
    Z=\int\D\phi_\pm \Z[\phi_\pm],\quad \Z[\phi_\pm]=\int_{\rho_0}\D\psi_\pm e^{iS_0[\psi_+,\phi_+]-iS_0[\psi_-,\phi_-]}\,.
\end{equation}
This can be interpreted as the generating functional of the slow modes $\phi_+,\phi_-$, which is obtained by treating $\phi_\a$ as the background field coupled to $\psi_\a$. Tracing back to the operator formalism, it reads
\begin{equation}
    \Z[\phi_\pm]=\Tr\big[U_{\rm fast}(\phi_+)\rho_0U_{\rm fast}^\hc(\phi_-)\big]\,.
\end{equation}

We generally consider a mixed state $\rho_0=\sum_n c_n|\Psi_n\ket\bra \Psi_n|$, and insert the complete basis with fixed background, $1=\int \D\psi^0|\psi^0,\phi_\a^0\ket\bra\psi^0,\phi_\a^0|$. One can verify that
\begin{equation}\label{eq:app_A=UU}
    \Z[\phi_\pm]=\sum_n c_n\big< \Psi_{n}(\phi_-^0)\big|U_{\rm fast}^\hc(\phi_-)U_{\rm fast}(\phi_+)\big|\Psi_{n}(\phi_+^0)\big>,
\end{equation}
where
\begin{equation}
    |\Psi_{n}(\phi_\a^0)\ket\equiv \int \D\psi^0 \bra \psi^0,\phi_\a^0|\Psi_n\ket|\psi^0,\phi_\a^0\ket
\end{equation}
are two states with norm
\begin{equation}
    \bra \Psi_{n}(\phi_\a^0)|\Psi_{n}(\phi_\a^0)\ket=\int \D\psi^0 \big|\Psi_n[\psi^0,\phi_\a^0]\big|^2
    \equiv f_n^2[\phi_\a^0],
\end{equation}
with real functional $f_n$. The normalization of the initial density matrix $\rho_0$ implies
\begin{equation}\label{eq:app_norm}
    \sum_n c_n f_n^2[\phi]=1,\quad \forall\, \phi.
\end{equation}

Motivated by Eq.~(\ref{eq:app_Z=intDA}), we divide the dependence of $\Z[\phi_\pm]$ into two parts: the effective initial state that depends on the boundary fields, and the effective action that depends on the bulk fields, which is
\begin{equation}\label{eq:app_A=rhoeiS}
    \Z[\phi_+,\phi_-]=\rho_{\rm eff}[\phi_+^0,\phi_-^0]\,e^{iS[\phi_+,\phi_-]},\quad \rho_{\rm eff}[\phi_+^0,\phi_-^0]\equiv \sum_n c_n f_n[\phi_+^0]f_n[\phi_-^0].
\end{equation}
The normalization condition~\eqref{eq:app_norm} implies $\rho_{\rm eff}[\phi^0,\phi^0]=1$.

Based on Eq.~\eqref{eq:app_A=UU} and \eqref{eq:app_A=rhoeiS}, several constraints for the effective action emerge:
\begin{itemize}
    \item By taking $\phi_+=\phi_-=\phi$ together with their initial configurations, Eq.~(\ref{eq:app_A=rhoeiS}) turns to
    \begin{equation}
        e^{iS[\phi,\phi]}=\Z[\phi,\phi]=\sum_n c_n\bra \Psi_{n}(\phi^0)|\Psi_{n}(\phi^0)\ket=1,
    \end{equation}
    which indicates $S[\phi,\phi]=0$.
    \item Taking the complex conjugation of $\Z[\phi_\pm]$, one obtains
    \begin{equation}
        \rho_{\rm eff}\,e^{-iS^*[\phi_+,\phi_-]}=\Tr\big[U_{\rm fast}(\phi_-)\rho_0 U_{\rm fast}^\hc(\phi_+)\big]=\rho_{\rm eff}\,e^{iS[\phi_-,\phi_+]},
    \end{equation}
    leading to $S^*[\phi_+,\phi_-]=-S[\phi_-,\phi_+]$.
    \item Cauchy-Schwarz inequality implies
    \begin{equation}
        \big|e^{iS}\big|\leq \frac{\sum_n c_n\| U_{\rm fast}(\phi_-)|\Psi_{n}(\phi_-^0)\ket \|\| U_{\rm fast}(\phi_+)|\Psi_{n}(\phi_+^0)\ket \|}{\sum_n c_nf_n[\phi_+^0]f_n[\phi_-^0]}=1.
    \end{equation}
    Hence, the imaginary part of the effective action should not be negative, $\mathrm{Im}S\geq 0$.
\end{itemize}
In the Keldysh basis, previous constraints become the form in the main text.

\section{Duality condition from contour permutation symmetry}\label{app:duality}

In the main text, we showed that the action of time reversal as complex conjugation on a transition amplitude is equivalent to exchanging the initial and final states, $|O\rangle$ and $|I\rangle$, and taking the Hermitian conjugate of the internal operators. For example, for the averaged partition function,
\begin{equation}\label{appeq:Z_TR}
    \overline Z\to \overline Z^*= \overline{\big< O\big|(\mathscr{T}e^{-i\int \sum_qJ_q(s)h_{Q,q}\,ds})^\hc\big|I\big>}_{J}=\overline{\big< O\big|\mathscr{T}e^{i\int \sum_qJ_q(-s)h_{Q,q}\,ds}\big|I\big>}_J,
\end{equation}
where the complete expression for the coupling average $\bra ...\ket_J$ is provided in the main text. We also note that introducing a permutation of the contour indices can transform the initial and final states back to their original positions. In this appendix, we present the details of this permutation.

Before proceeding, it is useful to express the operators $\mathcal N$, $\Phi$, and $\Phi^\dagger$ defined in the main text in terms of the generators of the $su(2)\oplus su(2)$ algebra. We define the Hermitian bosonic operators
\begin{eqnarray}
    Q^{\a\b}\equiv \sum_x Q_x^{\a\b}=\sum_x \sum_{j}i\psi^{\a}_{x,j}\psi^{\b}_{x,j},
\end{eqnarray}
where we denote $(\psi^{1},\psi^2,\psi^3,\psi^4)=(\psi^{L+},\psi^{R+},\psi^{L-},\psi^{R-})$. One can subsequently check that the densities corresponding to the generators of the emergent local $su(2)\oplus su(2)$ Lie algebra are
\begin{eqnarray}
    J_{\sigma,x}^x=-\frac{Q^{14}_x+\theta_{\sigma}Q^{23}_x}{2},\quad J_{\sigma,x}^y=-\frac{Q^{13}_x+\theta_{\sigma}Q^{42}_x}{2},\quad J_{\sigma,x}^z=\frac{Q^{12}_x+\theta_{\sigma}Q^{34}_x}{2},
\end{eqnarray}
for $\sigma=s,w$, where $\theta_\sigma=2\delta_{\sigma s}-1$. Their relations to the charge operator and ladder operators discussed in the main text are
\begin{equation}\label{eq:PhiJ}
    \Phi_{\sigma,x}^\hc=J^x_{\sigma,x}+iJ_{\sigma,x}^y,\quad \Phi_{\sigma,x}=J^x_{\sigma,x}-iJ_{\sigma,x}^y;\quad J^z_{s,x}=\N_{s,x}-\frac{N}{2},\quad J^z_{w,x}=\frac{\N_{w,x}}{2}.
\end{equation}
By substituting the local $su(2)\oplus su(2)$ algebra, $[J^j_{\sigma,x},J_{\rho,y}^k]=i\delta_{\sigma\rho}\delta_{xy}\epsilon^{jkl}J^l_{\sigma,x}$, it is easy to verify Eq.~(\ref{eq:su2su2}) in the main text. Since the final field-theory degrees of freedom do not contain fields from the weak sector, in the following discussion we keep only the operators and fields in the strong sector and suppress the index $s$. Using the global generators defined above, $J^i=\sum_x J_x^i$, the initial state $|O\rangle$ and final state $|I\rangle$ can be related as
\begin{equation}\label{appeq:|O>=|I>}
    |I\ket=e^{i\Phi^\dagger}|O\ket=e^{i(J^x+iJ^y)}|O\ket,\quad |O\ket=2^{-MN}e^{-i(J^x-iJ^z)}|I\ket.
\end{equation}
The states are normalized as $\langle O|O\rangle=1$ and $\langle I|I\rangle=2^{MN}$, where $M$ is the number of spatial lattice sites in the system.

Motivated by exchanging the states $|O\rangle$ and $|I\rangle$ back, we consider a permutation acting on the four contour indices, such that the boundary conditions of $|I\rangle$ are transformed into those of $|O\rangle$, and vice versa. The direction of this permutation is arbitrary, and we choose the ``counterclockwise'' direction
\begin{equation}
    \M\begin{Bmatrix}
        \psi^1 &\;&\psi^2\\\;&\;&\;\\\psi^3&\;&\psi^4
    \end{Bmatrix}\M^{-1}=\begin{Bmatrix}
        s_2\psi^2 &\;&s_4\psi^4\\\;&\;&\;\\s_1\psi^1&\;&s_3\psi^3
    \end{Bmatrix}.
\end{equation}
where $s_i$ are coefficients to be determined. The permutation acts on each $\psi$ operator, and the braces are introduced only for notational convenience, \textit{without} any matrix meaning.

Inserting $\M$ into the boundary conditions of $|O\ket$ and $|I\ket$, we obtain
\begin{equation}
\begin{split}
    s_2\big(\psi^2+is_2^{-1}s_4\psi^4\big)\M|O\ket&=s_1\big(\psi^1+is_1^{-1}s_3\psi^3\big)\M|O\ket=0;\\
    is_1\big(\psi^1+is_1^{-1}s_2\psi^2\big)\M|I\ket&=is_3\big(\psi^3-is_3^{-1}s_4\psi^4\big)\M|I\ket=0.
\end{split} 
\end{equation}
Since we suppose that $\M|O\ket\propto|I\ket$ and $\M|I\ket\propto|O\ket$, it enforces $s_1=s_2=s_4=-s_3\equiv s$. We can act with $\M$ on Eq.~\eqref{appeq:|O>=|I>}, and further determine $s$ by imposing self-consistency. To begin with, the permutation acts on three generators of the strong sector as
\begin{equation}\label{appeq:MJM}
    \M J^x\M^{-1}=-s^2 J^x,\quad \M J^y\M^{-1}=s^2 J^z,\quad \M J^z\M^{-1}=s^2 J^y.
\end{equation}
Hence, acting $\M$ on Eq.~\eqref{appeq:|O>=|I>} gives
\begin{equation}
\begin{split}
    \M|I\ket&=\M e^{i(J^x+iJ^y)}\M^{-1}\M|O\ket=e^{-is^2(J^x-iJ^z)}\M|O\ket,\\
    \M|O\ket&=\M e^{-i(J^x-iJ^z)}\M^{-1}\M|I\ket=e^{is^2(J^x+iJ^y)}\M|I\ket.
\end{split}
\end{equation}
Comparing with Eq.~\eqref{appeq:|O>=|I>} itself, one obtains a self-consistent transformation relation when $s=\pm 1$:
\begin{equation}
    \M|O\ket=2^{-MN/2}|I\ket,\quad \M|I\ket=2^{MN/2}|O\ket,
\end{equation}
where the prefactors of the transformed states are chosen so that $\mathcal M^{-1}=\mathcal M^\dagger$. The choice $s=1$ or $-1$ is immaterial, and in the main text we take $s=1$.

Under the contour permutation $\mathcal M$, the component of the Hamiltonian that evolves states in the quadrupled Hilbert space, $H_Q(t)=\sum J_q(t)h_{Q,q}$, with
\begin{equation}
\begin{split}
    h_{Q,q}&\equiv h_{D,+;q}-(-1)^{q/2}h_{D,-;q}\\
    &=h_{L+,q}-(-1)^{q/2}h_{R+,q}-(-1)^{q/2}h_{L-,q}+h_{R-,q}\,,
\end{split}
\end{equation}
transforms as $\M\, h_{Q,q}\,\M^{-1}=-(-1)^{q/2}h_{Q,q}$. Therefore, under the permutation transformation, the averaged partition function \eqref{appeq:Z_TR} becomes
\begin{equation}
    \overline Z\to(\M\T)\overline{Z}(\M\T)^{-1}=\overline{\big< I\big|\mathscr{T}e^{-i\int \sum_q [(-1)^{q/2}J_q(-s)]h_qds}\big|O\big>}_J=\overline Z.
\end{equation}
Consequently, the effective action satisfies a duality condition
\begin{equation}\label{eq:app,S[n,phi]}
    S[n(t),\phi(t)]=S\big[\widetilde{n}(-t),\widetilde{\phi}(-t)\big],
\end{equation}
where $\widetilde n$ and $\widetilde\phi$ are the field transformations corresponding to the operators associated with the field degrees of freedom under time reversal and the permutation transformation.

To obtain the transformations of these field degrees of freedom, we trace them back to the transformations of the $su(2)$ generator operators. As stated at the beginning of this section, time reversal acts on operators effectively by Hermitian conjugation. It therefore leaves these Hermitian generators unchanged, but reverses the time ordering, leading to $t\to -t$. On the other hand, the effect of the permutation is given in Eq.~\eqref{appeq:MJM} with $s=1$. The combined effect is thus
\begin{equation}
    J^z(t)\xrightarrow{\,\M\T\,}J^y(-t),\quad J^y(t)\xrightarrow{\,\M\T\,}J^z(-t).
\end{equation}
In the operator notation used in the main text, the last two transformations are written as
\begin{equation}
    n-\frac{1}{2}\xleftrightarrow{\; \mathcal{MT} \;}\frac{i}{2}\big(\varphi-\varphi^\hc\big),
\end{equation}
where the time arguments before and after the transformation are $t$ and $-t$, respectively, and are suppressed here. Substituting further the expansion of the field degrees of freedom in the path integral, Eq.~\eqref{eq:delta_n_delta_phi} in the main text, we finally obtain
\begin{equation}
    n\xleftrightarrow{\; \mathcal{MT} \;} in\sin\phi-\frac{1}{2}\left(e^{i\phi}-1\right),
\end{equation}
again with time arguments $t$ before the transformation and $-t$ after the transformation.

\section{Expressing OTOC in terms of EFT degrees of freedom}
\label{sec:otoc_correlator}

In this appendix, we present the detailed derivation of the representation of the Majorana OTOC in terms of the degrees of freedom of the effective field theory (EFT). We first note the two-point correlation function expressed in terms of the bosonic operators in the main text
\begin{equation}\label{appeq:<nPhi>}
    \C_m(x,t)=i\left\bra n_x(t)\left(\varphi_0^\hc(0)-\varphi_0(0)^m\right)\right\ket=\frac{i}{N^{m+1}}\left\bra \N_x(t)\left(\Phi_0^\hc(0)-\Phi_0(0)^m\right)\right\ket.
\end{equation}
To show that this two-point correlation function is equal to the OTOC, we expand \(\mathcal N\), \(\Phi\), and \(\Phi^\dagger\) in terms of Majorana operators. For clarity, we present the derivation for \(m=1\); the case \(m>1\) follows in a completely analogous manner. For \(m=1\), the above correlation function expands as follows:
\begin{equation}
    \C_1(x,t)=\frac{i}{2N^2}\sum_{jk}\Big< I\Big|\mathscr{C}\big\{\big(1+i\psi^1_{x,j}(t)\psi_{x,j}^2(t)+i\psi_{x,j}^3(t)\psi_{x,j}^4(t)\big)\big(\psi_{0,k}^1\psi_{0,k}^3-\psi^2_{0,k}\psi_{0,k}^4\big)\big\}\Big|O\Big>,
\end{equation}
where we relabel $(\psi^1,\psi^2,\psi^3,\psi^4)$ as $(\psi^{L+},\psi^{R+},\psi^{L-},\psi^{R-})$. In this expression, $\mathscr{C}$ represents the ordering along the four-branch contour, which arranges operators as
\begin{equation}
    \psi^4(0)\gets\psi^3(0)\gets\psi^4(t)\gets \psi^3(t)\gets \psi^2(t)\gets \psi^1(t)\gets \psi^2(0)\gets \psi^1(0).
\end{equation}
Therefore, we impose the ordering $\mathscr{C}$ and, unless explicitly required, omit the spatial dependence as well as the summation over the flavor indices $j,k$. Under these conventions, the correlator takes the form
\begin{equation}
    =\frac{i}{2N^2}\big< I\big| \psi^{4}_0\psi^{2}_0-\psi^{3}_0\psi^{1}_0 +i\psi^{3}_0\psi^{2}_t\psi^{1}_t\psi^{1}_0
    -i\psi^{4}_0\psi^{2}_t\psi^{1}_t\psi^{2}_0
    +i\psi^{3}_0\psi^{4}_t\psi^{3}_t\psi^{1}_0
    -i\psi^{4}_0\psi^{4}_t\psi^{3}_t\psi^{2}_0\big|O\big>.
\end{equation}
We then consider that $H_Q|I\ket=H_Q|O\ket=0$ such that
\begin{equation}
    =\frac{i}{2N^2}\big< I\big| \psi^{4}_0\psi^{2}_0-\psi^{3}_0\psi^{1}_0+i\psi^{3}_t\psi^{2}_0\psi^{1}_0\psi^{1}_t
    -i\psi^{4}_t\psi^{2}_0\psi^{1}_0\psi^{2}_t
    +i\psi^{3}_t\psi^{4}_0\psi^{3}_0\psi^{1}_t
    -i\psi^{4}_t\psi^{4}_0\psi^{3}_0\psi^{2}_t\big|O\big>.
\end{equation}
From the boundary condition of $\bra I|$, the first two terms evaluate to $\bra I|\psi^{a+2}\psi^a|O\ket=(-1)^{a+1}i/2$. Using the anticommutation relation $\{\psi^a,\psi^b\}=\delta^{ab}$, this leads to
\begin{equation}
\begin{split}
    \C_1(x,t)&=\frac{1}{2}-\frac{1}{2N^2}\big< I\big|\psi_t^3\psi_0^1\psi_t^1\psi_0^2+\psi_t^4\psi_0^2\psi_t^2\psi_0^1+\psi_t^1\psi_t^3\psi_0^3\psi^4_0+\psi_t^2\psi_t^4\psi_0^4\psi_0^3\big|O\big>\\
    &=\frac{1}{2}+\frac{1}{2N^2}\big< I\big| \psi^1_t\psi^1_0\psi^1_t\psi^1_0+\psi^2_t\psi^2_0\psi^2_t\psi^2_0-\psi^3_t\psi^3_t\psi^3_0\psi^3_0-\psi^4_t\psi^4_t\psi^4_0\psi^4_0\big|O\big>\\
    &=\frac{1}{4}+\frac{1}{N^2}\sum_{jk}\Tr\big[\psi_{x,j}(t)\psi_{0,k}\psi_{x,j}(t)\psi_{0,k}\big]=\frac{1}{2N^2}\sum_{jk}\tr\big[ |\{\psi_{x,j}(t),\psi_{0,k}\}|^2\big]
\end{split}
\end{equation}
The second equality is obtained by substituting the boundary conditions of $\bra I|$ and $|O\ket$. Hence, the correlator $\C_{1}(x,t)$ captures the dynamical behavior of the OTOC.

Once the two-point correlation function \eqref{appeq:<nPhi>} is shown to be equal to the OTOC, it can be represented in the path-integral formulation in terms of the EFT degrees of freedom as
\begin{equation}\label{appeq:Cm(t)}
    \C_m(t)=\Big< n(x,t)\Big(e^{i\phi(0,0)}-n(0,0)\big(e^{i\phi(0,0)}-e^{-i\phi(0,0)}\big)\Big)^m\Big>,
\end{equation}
where we substitute Eq.~\eqref{eq:delta_n_delta_phi}. Expanding the above correlation function gives the correlation with $p=0$ order of $n(0,0)$, together with higher-$p$ orders of $n(0,0)$ correlation functions. We now show that the latter are suppressed in the large-$N$ limit. Considering the $p$-th order of $n(0,0)$ term and evaluating the path integral in the saddle-point approximation, we obtain
\begin{equation}
    \begin{split}
        \C_m\supset & \frac{N}{Z}\int\D n_0\D\phi_0\, n_*(n_0;x,t)n^p(0,0)\big(e^{i\phi_0(0)}-e^{-i\phi_0(0)}\big)^pe^{i(m-p)\phi_0(0)}e^{-iN\int dx\, n_0\phi_0}\\
        =&\int \D n_0\, n_*(n_0;x,t)n^p(0,0) \sum_{r=0}^p \binom{p}{r} (-1)^r \delta\Big(n_0(x)-\frac{\delta(x)}{N}(m-2r)\Big)\\
        =& \Big(\frac{\delta(0)}{N}\Big)^p\cdot \sum_{r=0}^{\min [m/2,p]}(-1)^r\binom{p}{r} \,n_*\Big(\frac{\delta(x)}{N}(m-2r);x,t\Big)(m-2r)^p\,.
    \end{split}
\end{equation}
Returning to the actual discrete model, the divergent $\delta(0)$ is in fact the inverse lattice spacing, namely $\delta(0)\sim 1/\Delta x$. The large-$N$ limit of the $(1+1)d$ model requires $N\Delta x\gg 1$. Therefore, the $p$-th order terms in $n(0,0)$ are suppressed by $(N\Delta x)^{-p}$, and only the $p=0$ term dominates Eq.~\eqref{appeq:Cm(t)}. Thus
\begin{equation}
    \C_m(t)=\left\bra n(x,t)e^{im\phi(0,0)}\right\ket\left[1+O\left((N\Delta x)^{-1}\right)\right].
\end{equation}

\section{Brownian SYK model}\label{appsec:SYK}

In this appendix, we present the explicit calculation of the effective field theory for the Brownian SYK model. Starting from the Hamiltonian defined in the main text and integrating out the Brownian random couplings, the microscopic action can be written in the complex basis as $S[c^{\a}]$, with degrees of freedom denoted as $(c^1,c^2,c^3,c^4)\equiv (c^{+\hc},c^+,c^{-\hc},c^-)$. Introducing the bosonic bilocal fields $G^{\alpha\beta}$ and the corresponding self-energy fields $\Sigma^{\alpha\beta}$ as Lagrange multipliers, one obtains the action
\begin{equation}\label{appeq:S[EG]}
    \begin{split}
        \frac{iS[\Sigma,G]}{N}=&\log \Pf\,\left[\frac{1}{2}\sigma^0\otimes \sigma^x(\partial_t-2i\V\sin{k})+\Sigma\right]\\
        &+\sum_{\a,\b\neq\a}\sum_x\int_{t_1t_2} G^{\a\b}_x\Sigma^{\a\b}_x-\left(\frac{\J_2}{4} G^{\a\b}_x G^{\bar{\a}\bar{\b}}_{x+1}
        +\frac{\J_4}{8} F^{\a\b}_x(G)\right)\delta(t_{12}),
    \end{split}
\end{equation}
where $\bar 1=2$, $\bar 3=4$ and vice versa. Here we suppress the time-dependence of the bilocal fields, and the last function reads
\begin{equation}
\begin{split}
    F^{\a\b}(G)=\sum_{y}\Delta_{x,y}&\left[(2\delta_{\a\bar\b}-1) (G_x^{\a\b})^2G_{y}^{\a\b}G_{y}^{\bar\a\bar\b}\right.\\
    &\left.-(1-\delta_{\a\bar\b})\left((G_x^{\a\b})^2G_{y}^{\bar\a\b}G_{y}^{\a\bar\b}+2G_x^{\a\b}G_x^{\a\bar\b}G_{y}^{\a\b}G_{y}^{\bar\a\b}\right)\right],
\end{split}
\end{equation}
where $\Delta_{x,y}=\delta_{y,x+1}+\delta_{y,x-1}$ depends on the specific hopping structure of the model.

Since the action satisfies $S[\Sigma,G]\sim N$, its treatment in the large-$N$ limit proceeds in three steps. First, we take the mean-field approximation to obtain the saddle-point equations and their solutions. We then consider fluctuations around the saddle point, derive the fluctuation action up to the quadratic order, and integrate out the gapped modes. Finally, we expand the local interaction terms beyond the quadratic order in order to compare with the minimal effective action obtained in the general theory.

\subsection{Mean-field solution}

The mean-field approximation neglects the spatial dependence of the fields, so that the path integral is dominated by the uniform saddle-point. Varying with respect to the $G$- and $\Sigma$-fields gives the Schwinger-Dyson equations
\begin{equation}\label{appeq:E=G}
    \overline{\Sigma}^{\a\b}(\omega)=\frac{\J_2}{2} \overline{G}^{\bar{\a}\bar{\b}}_0+\frac{\J_4}{4} (2\delta_{\a\bar\b}-1)\left[\overline{G}^{\bar{\a}\bar{\b}}_0+3\overline{G}^{\bar{\a}\bar{\b}}_0\left(\overline{G}^{{\a}{\b}}_0\right)^2\right],
\end{equation}
\begin{equation}\label{appeq:G=E}
    \overline{G}(\omega)=\int \frac{dk}{2\pi}\left(2\overline{\Sigma}-i\sigma^0\otimes\sigma^x(\omega+2\V\sin{k})\right)^{-1}\equiv\int_k \overline{G}_k(\omega),
\end{equation}
for $\a\neq \b$. Here, we simply denote $\bar{G}^{\a\b}(t=0)$ as $\bar{G}_0^{\a\b}$. Its different matrix elements are related by the boundary conditions \eqref{eq:psipsi|0>} and \eqref{eq:psiLpsiR|I>=0}, which close the system of equations.

In fact, when imposing the boundary conditions, one must carefully distinguish $t=0$ from $t\to 0^+$. The main origin of this distinction is that the Majorana fields in the path integral are treated as Grassmann numbers rather than operators. The boundary conditions at $t\to 0^+$ impose:
\begin{equation}
    \overline{G}^{12}_{0^+}=\overline{G}_{0^+}^{34}=i\overline{G}^{24}_{0^+}\equiv p,\quad \overline{G}^{21}_{0^+}=\overline{G}_{0^+}^{43}=i\overline{G}^{31}_{0^+}\equiv iq,
\end{equation}
with others vanishing. However, at $t=0$, the Grassmann nature of the fields implies that $\overline G^{\alpha\bar\alpha}_0\neq \overline G^{\alpha\bar\alpha}_{0^+}$. The boundary condition is therefore taken to be
\begin{equation}
    \overline G^{12}_{0}=\overline G^{34}_{0}=-\overline G^{21}_{0}=-\overline G^{43}_{0}\equiv r.
\end{equation}
One can verify that the self-consistent solution satisfying the Schwinger-Dyson equations and the boundary conditions is obtained at $r=-1/2$, $p=0$ and $q=-i$, and is given by
\begin{equation}
    \overline{G}(t)=e^{-\Gamma|t|}J_0(2\V|t|)\begin{pmatrix}
        0 & -\theta(-t)& -i & 0\\\theta(t) & 0&0&0\\
        i&0&0&-\theta(-t)\\0&0&\theta(t)&0
    \end{pmatrix},
\end{equation}
where $\Gamma=\J_2/2+\J_4/4$ and $J_0(x)$ is the Bessel function.

\subsection{Saddle-point expansion}

After obtaining the mean-field, or saddle-point, solution, we expand the fields around the mean-field background as $G^{\a\b}_x(t_1,t_2)=\overline{G}^{\a\b}(t_{12})+\delta G_x^{\a\b}(t_1,t_2)$ and $\Sigma^{\a\b}_x(t_1,t_2)=\overline{\Sigma}^{\a\b}(t_{12})+\delta \Sigma_x^{\a\b}(t_1)\delta(t_{12})$, where the self-energy fluctuations are diagonal in time, because the interaction is strictly time-local.

Substituting the fluctuation expansion into the effective action, one can in principle keep terms to arbitrary order in the fluctuations. However, in order to integrate out $\delta\Sigma$ analytically, it is convenient to retain only the quadratic fluctuation terms of $\delta\Sigma$, which arise from the kernel term
\begin{equation}\label{appeq:kernel_calcu}
\begin{split}
    \frac{iS_{\rm eff}}{N}\supset& \log\Pf\left[-\frac{i}{2}\sigma^0\otimes \sigma^x(\omega+2\V\sin{k})+\overline\Sigma+\delta\Sigma\right]\\
    \supset&\frac{1}{2}\Tr\log(1+\overline G_k\delta\Sigma)\supset-\int_{k\omega}\delta\Sigma(k,\omega)M_{k,\omega}\delta\Sigma^T(-k,-\omega),
\end{split}
\end{equation}
where $\delta\Sigma=\delta(\Sigma^{12},\Sigma^{13},\Sigma^{14},\Sigma^{23},\Sigma^{24},\Sigma^{34})$, and
\begin{equation}
\begin{split}
    M_{k,\omega}=\frac{2}{|\Omega_{k,\omega}|^2}\begin{pmatrix}
        0& -\Omega_{k,\omega}^* &0&0&0&0\\ \Omega_{k,\omega} & i(\Omega_{k,\omega}-\Omega_{k,\omega}^*) &0&0&-i\Omega_{k,\omega}&\Omega_{k,\omega}\\0&0&0&0&0&0\\
        0&0&0&0&0&0\\0&i\Omega_{k,\omega}^*&0&0&0&0\\0&-\Omega_{k,\omega}^*&0&0&0&0
    \end{pmatrix},
\end{split}
\end{equation}
$$\Omega_{k,\omega}=\big((\omega-2i\Gamma)^2-16\V^2\sin^2(k/2)\big)^{1/2}.$$
Here, $M_{k,\omega}$ is a rank-2 matrix. To reduce its dimension, we introduce a transformation matrix $P$ such that
\begin{equation}
    P^TM_{k,\omega}P=\begin{pmatrix}
        m_{k,\omega} &0_{2\times4}\\0_{4\times2}&0_{4\times4}
    \end{pmatrix},\qquad m_{k,\omega}=\frac{1}{|\Omega_{k,\omega}|^2}\begin{pmatrix}
        0& -\Omega_{k,\omega}^*\\\Omega_{k,\omega} & i(\Omega_{k,\omega}-\Omega_{k,\omega}^*)
    \end{pmatrix}.
\end{equation}
At the same time, $\delta\Sigma$ and $\delta G=\delta(G^{12},G^{13},G^{14},G^{23},G^{24},G^{34})$ are transformed as $\delta\tilde \Sigma= P^{-1}\delta\Sigma$ and $\delta\tilde G=P^T\delta G$.

Expanding the second term of the action \eqref{appeq:S[EG]} gives the coupling between \(\delta G\) and \(\delta\Sigma\). Integrating out \(\delta\Sigma\) therefore reduces the number of degrees of freedom to \(\mathrm{rank}(m_{k,\omega})=2\), and imposes the constraints
\begin{equation}\label{appeq:deltaG_constraint}
    \delta G^{12}_x=\delta G^{34}_x=i\delta G_x^{24},\quad \delta G^{14}_x=\delta G^{23}_x=0.
\end{equation}
Substituting the constraint, the first two terms of Eq.~\eqref{appeq:S[EG]} contribute the kernel term of the effective action
\begin{equation}\label{appeq:S_eff_kernel}
    \frac{iS_{\rm eff}}{N}\supset \int_{k,\omega}\delta G^{12}_{-k}(-\omega)\cdot\Omega_{k,\omega}\cdot[\delta G^{13}_k(\omega)-i\delta G^{12}_k(\omega)].
\end{equation}

Finally, we consider the interaction term. Keeping \(\delta G^{24}\) explicitly and substituting the other constraints on the fluctuations, we obtain
\begin{equation}\label{appeq:int}
\begin{split}
    \frac{iS_{\rm eff}}{N}\supset-\int_{x,t}\sum_y\Delta_{x,y}&\left\{\frac{\J_2}{2}(-\delta G^{12}_x\delta G^{12}_{y}+\delta G^{13}_x\delta G^{24}_{y})\right.\\
    &+\frac{\J_4}{4}\left[(\delta G_x^{12})^2+(1-\delta G^{12}_x)^2\right]\delta G_y^{12}(1-\delta G^{12}_y)\\
    &+\left.\frac{\J_4}{4}\left[(1+i\delta G^{13}_x)^2+(-i\delta G_x^{24})^2\right](1+i\delta G^{13}_y)(-i\delta G_y^{24})\right\}.
\end{split}
\end{equation}
To quadratic order, the interaction term gives an additional contribution to the effective action
\begin{equation}\label{appeq:mass_quadratic}
    \frac{iS_{\rm eff}}{N}\supset -\left(2\Gamma+\J_4\right)\int_{k,\omega}\cos{k}\,\left[\delta G^{24}_{-k}(-\omega)\delta G^{13}_k(\omega)-|\delta G^{12}_k(\omega)|^2\right].
\end{equation}
Here, $\delta G^{24}$ is kept explicitly for convenience in deriving the higher-order terms in the next subsection, where the relation $\delta G^{12}=i\delta G^{24}$ is no longer exact. However, for obtaining the quadratic effective action this constraint can be safely imposed, yielding the quadratic effective action
\begin{equation}\label{appeq:full_quadratic}
    \frac{S_{\rm eff}}{N}= \int_{k,\omega}\delta G^{12}_{-k}(-\omega)\cdot[-i\Omega_{k,\omega}+(2\Gamma+\J_4)\cos{k}]\cdot[\delta G^{13}_k(\omega)-i\delta G^{12}_k(\omega)].
\end{equation}

\subsection{Higher-order terms in the effective action}

We now consider higher-order terms in the fluctuation expansion. To this end, we first relate the bilocal field $\delta G$ to the degrees of freedom used to construct the EFT in the main text. Specifically,
\begin{equation}\label{appeq:deltaG_nphi}
    \delta G^{12}=\delta G^{34}=n,\quad \delta G^{13}=\delta\vphi=-i(1-n)e^{i\phi}+i,\quad \delta G^{24}=-\delta\vphi^\hc=-in e^{-i\phi}.
\end{equation}
Here we keep only the degrees of freedom carrying the $s$-index. The $w$-sector degrees of freedom are absent from the general EFT and, moreover, the corresponding fluctuations of bilocal field are indeed constrained to vanish by Eq.~\eqref{appeq:deltaG_constraint}.

At leading order, one has $\delta G^{12}=i\delta G^{24}$, in agreement with the constraint \eqref{appeq:deltaG_constraint}. Higher-order fluctuations, however, deviate from Eq.~\eqref{appeq:deltaG_constraint}, which due to the fact that $\delta\Sigma$ has been expanded only up to quadratic order.

We argue this point qualitatively. If the expansion in Eq.~\eqref{appeq:kernel_calcu} is kept to fourth or even higher order, it generates a perturbation to $M_{k,\omega}$ and changes its rank. For instance, if the effective perturbation from the higher-order terms is a diagonal matrix $\epsilon \mathbb{I}$, then $\mathrm{rank}(M_{k,\omega}+\epsilon \mathbb{I})=5$. In this case, the action for $(\delta G^{14},\delta G^{23})$ can still be separated out, or even integrated out, while the remaining variables are subject only to the constraint $\delta G^{12}=\delta G^{34}$, with no additional constraint on $\delta G^{24}$. Therefore, by comparison with Eq.~\eqref{appeq:mass_quadratic}, the reasonable form of the quadratic effective action in Eq.~\eqref{appeq:full_quadratic} is modified to
\begin{equation}
    \frac{S_{\rm eff}}{N}= \int_{k,\omega}[\Omega_{k,\omega}+(2i\Gamma+i\J_4)\cos{k}]\,\left[\delta G^{24}_{-k}(-\omega)\delta G^{13}_k(\omega)-|\delta G^{12}_k(\omega)|^2\right].
\end{equation}
The rationale for this modification is twofold. First, at quadratic order it reduces back to Eq.~\eqref{appeq:full_quadratic}. Second, it prevents the $\J_2$ interaction from generating a mass term, as required by the emergent U(1) symmetry of the $q=2$ theory.

To obtain the higher-order terms in the effective action, we expand the interaction term in Eq.~\eqref{appeq:int} to arbitrary order in the fluctuations and substitute the full fluctuation form in Eq.~\eqref{appeq:deltaG_nphi}. For the higher-order terms, we keep only the relevant on-site contributions and finally expand to quadratic order in $\phi$. The quadratic interaction term is obtained as \eqref{appeq:mass_quadratic}. The on-site cubic fluctuation term is obtained from Eq.~\eqref{appeq:int} by taking $\Delta_{x,y}=2\delta_{xy}$, which gives
\begin{equation}
    \frac{S_{\rm eff}}{N}\supset \frac{i\J_4}{2}\int_{x,t}4(\delta G^{12}_x)^3+3i(\delta G^{13}_x)^2\delta G^{24}_x+i(\delta G^{24}_x)^3.
\end{equation}
Similarly, the on-site quartic fluctuation term is
\begin{equation}
    \frac{S_{\rm eff}}{N}\supset-\frac{i\J_4}{2}\int_{x,t}2(\delta G^{12}_x)^4+(\delta G^{13}_x)^3\delta G^{24}_x+\delta G^{13}_x(\delta G^{24}_x)^3.
\end{equation}
Substituting Eq.~\eqref{appeq:deltaG_nphi} and retaining terms up to order $\phi^2$, we finally obtain the complete effective action:
\begin{equation}
    \frac{S_{\rm eff}}{N}=\int_{x,t} n(\partial_t+D\nabla^2_x)\phi+\lambda n(1-n)(1-2n)\phi+i\lambda n(1-n)(1-2n+2n^2)\phi^2+O(\phi^3),
\end{equation}
where the Lyapunov exponent is $\lambda=\mathcal J_4$, and the diffusion constant is $D=\Gamma(1+\V^2/\Gamma^2)+\J_4/2$. The analytic result for the effective action obtained by the above method reproduces the minimal action derived from the general EFT analysis.

\section{Scaling behavior of the OTOC front in the clean system}\label{appsec:clean}

In the main text, we pointed out that the clean model has a nonlocal effective action, while the OTOC can nevertheless be expressed as a Fourier transform
\begin{equation}\label{appeq:J_int1}
    \C_m(x,t)=im\int_{k,\omega}\frac{e^{i(kx-\omega t)}}{\sqrt{\omega^2-16\V^2\sin^2(k/2)}}=\frac{m}{2}\int_k J_0\big(4\V t|\sin(k/2)|\big)e^{ikx}\,.
\end{equation}
Here the $\omega$ integral is taken over $[-\omega_k,\omega_k]$ with $\omega_k\equiv4\V\sin(k/2)$, because the relevant contribution comes from the branch-cut discontinuity. With the physical branch prescription, the square root on this cut satisfies $(\omega^2-\omega_k^2)^{1/2}=i(\omega_k^2-\omega^2)^{1/2}$, so that the full correlation function is real.

As before, we keep only the real part as the actual contribution of the integral to the correlation function. To evaluate the Fourier transform, we first prove the following identity:
\begin{equation}
    J_0\big(2\tau |\sin(k/2)|\big)=\sum_{\nu\in \mathbb{Z}}J_\nu(\tau)^2e^{i\nu k}.
\end{equation}

\begin{proof}
    We start from the integral representation of the zeroth Bessel function,
\begin{equation}
    J_0(|\mathbf{x}|)=\int_0^{2\pi} e^{i|\mathbf{x}|\cos \theta}\frac{d\theta}{2\pi}=\int_0^{2\pi} e^{i\mathbf{x}\cdot \mathbf{n}_\theta}\frac{d\theta}{2\pi}\,,
\end{equation}
where $\mathbf n_\theta$ is the unit vector making an angle $\theta$ with $\mathbf x$. Given two vectors of magnitude $\tau$, namely $\mathbf{a}=\tau(1,0)$ and $\mathbf{b}=\tau(\cos{k},\sin{k})$, the Bessel function $J_0(|\mathbf a-\mathbf b|)$ can, by construction, be written as
\begin{equation}
   J_0\big(2\tau|\sin(k/2)|\big)=J_0(|\mathbf a-\mathbf b|)=\int_0^{2\pi}e^{i\tau(\cos\theta-\cos{(\theta-k)})}\frac{d\theta}{2\pi}.
\end{equation}
To evaluate this integral, we apply the Jacobi-Anger expansion
\begin{equation}
    e^{i\tau \cos\theta}=\sum_{\nu\in \mathbb{Z}}i^\nu J_\nu(\tau)e^{i\nu\theta}.
\end{equation}
After integrating over $\theta$, which collapses one of the summation indices, the integral reduces to
\begin{equation}
    J_0(|\mathbf{a}-\mathbf{b}|)=\sum_{\nu\in \mathbb{Z}}(-1)^\nu J_\nu(\tau)J_{-\nu}(\tau)e^{i\nu k}=\sum_{\nu\in \mathbb{Z}}J_\nu(\tau)^2e^{i\nu k}.
\end{equation}
Thus the identity is proved.
\end{proof}

Using this identity, the Fourier transform in \eqref{appeq:J_int1} then becomes
\begin{equation}
    \C_m(x,t)=\frac{m}{2}J_x(2\V t)^2,\quad x\in\mathbb{Z}.
\end{equation}
From this explicit result, one can straightforwardly derive several asymptotic behaviors of the OTOC. In the late-time limit inside the light-cone, namely $\tau\equiv 2\mathcal V t>|x|\gg 1$, the asymptotic envelope of this function reproduces Eq.~\eqref{eq:sec6_tail} in the main text. Near the late-time wavefront, where $x,\tau\gg1$ while $x-\tau\ll \tau$, the asymptotic behavior of the Bessel function is nontrivial and reads
\begin{equation}
\begin{split}
    J_x(\tau)\equiv& \int_{-\pi}^{\pi} \frac{d\theta}{2\pi}e^{i(x\theta-\tau\sin\theta)}\approx \int_{-\pi}^{\pi}\frac{d\theta}{2\pi}e^{i(x-\tau)\theta+\frac{i}{3!}\tau \theta^3}\\
    =&(2/\t)^{1/3}\mathrm{Ai}\big[(2/\t)^{1/3}(x-\tau)\big],
\end{split} 
\end{equation}
where $\mathrm{Ai}(x)$ is the Airy function. Therefore, the asymptotic behavior of the OTOC is
\begin{equation}
    \C_m(x,t)\simeq \frac{m}{2}\Big(\frac{2}{\t}\Big)^{2/3}\mathrm{Ai}\Big[\Big(\frac{2}{\t}\Big)^{1/3}(x-\tau)\Big]^2
    \simeq  \frac{m}{4\pi\sqrt{2\tau(x-\tau)}}e^{-\frac{4\sqrt{2}}{3}\frac{(x-\tau)^{3/2}}{\tau^{1/2}}},
\end{equation}
This establishes the late-time asymptotic form of the OTOC wavefront in the clean model discussed in the main text, demonstrating the scaling behavior $(x - vt) \sim t^{1/3}$.

\bibliographystyle{jhep}
\bibliography{ref}

\end{document}